\documentclass[acmtog]{acmart}

\AtBeginDocument{%
  }

\setcopyright{licensedothergov}

\acmJournal{TOG}

\acmSubmissionID{685}

\usepackage{subcaption}
\usepackage{multirow}
\usepackage{xcolor,colortbl}
\usepackage{algorithm}
\usepackage{algpseudocode}
\usepackage{tikz}
\usepackage{graphicx}
\usetikzlibrary{spy,backgrounds}
\usepackage{soul}

\usepackage[percent]{overpic}

\newcommand{\TODO}[1]{}

\setcopyright{licensedothergov}
\acmYear{2023} \acmVolume{42} \acmNumber{4} \acmArticle{1} \acmMonth{8} \acmPrice{15.00} \acmDOI{10.1145/3592433}

\begin{document}

\title{3D Gaussian Splatting for Real-Time Radiance Field Rendering}

\author{Bernhard Kerbl}
\orcid{0000-0002-5168-8648}
\authornote{Both authors contributed equally to the paper.}
\email{bernhard.kerbl@inria.fr}
\affiliation{%
	\institution{Inria, Universit\'e C\^ote d'Azur}
	\country{France}
}
\author{Georgios Kopanas}
\orcid{0009-0002-5829-2192}
\authornotemark[1]
\email{georgios.kopanas@inria.fr}
\affiliation{%
	\institution{Inria, Universit\'e C\^ote d'Azur}
	\country{France}
}
\author{Thomas Leimk\"{u}hler}
\orcid{0009-0006-7784-7957}
\email{thomas.leimkuehler@mpi-inf.mpg.de}
\affiliation{%
	\institution{Max-Planck-Institut f\"{u}r Informatik}
	\country{Germany}
}
\author{George Drettakis}
\orcid{0000-0002-9254-4819}
\email{george.drettakis@inria.fr}
\affiliation{%
	\institution{Inria, Universit\'e C\^ote d'Azur}
	\country{France}
}

\def\Dg{DG}

\newcommand{\ADDITION}[1]{#1}
\newcommand{\REMOVAL}[1]{}
\newcommand{\CORRECTION}[2]{#2}

\begin{abstract}
Radiance Field methods have recently revolutionized novel-view synthesis of scenes captured with multiple photos or videos. However, achieving high visual quality still requires neural networks that are costly to train and render, while recent faster methods inevitably trade off speed for quality. For unbounded and complete scenes (rather than isolated objects) and 1080p resolution rendering, no current method can achieve real-time display rates. We introduce three key elements that allow us to achieve state-of-the-art visual quality while maintaining competitive training times and importantly allow high-quality real-time ($\geq30$~fps) novel-view synthesis at 1080p resolution. First, starting from sparse points produced during camera calibration, we represent the scene with 3D Gaussians %
that preserve desirable properties of continuous volumetric radiance fields for scene optimization while avoiding unnecessary computation in empty space; Second, we perform interleaved optimization/density control of the 3D Gaussians, notably optimizing anisotropic covariance to achieve an accurate representation of the scene; Third, we develop a fast visibility-aware rendering algorithm that supports anisotropic splatting and both accelerates training and allows real-time rendering. We demonstrate state-of-the-art visual quality and real-time rendering on several established datasets.
\end{abstract}



\begin{teaserfigure}
	\includegraphics[width=\textwidth]{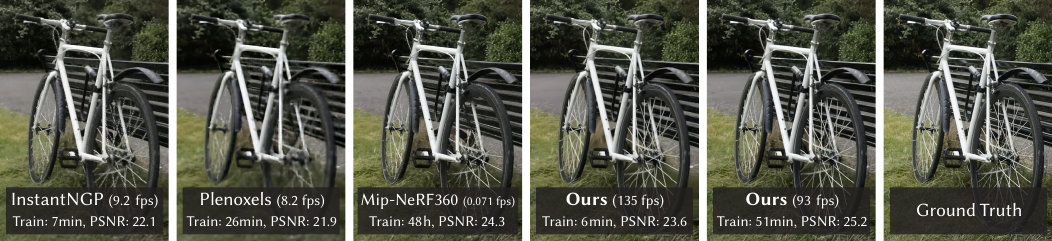}
	\caption{
	\label{fig:teaser}
    Our method achieves real-time rendering of radiance fields with quality that equals the previous method with the best quality ~\cite{barron2022mipnerf360}, while only requiring optimization times competitive with the fastest previous methods~\cite{plenoxels,mueller2022instant}.
    Key to this performance is a novel 3D Gaussian scene representation coupled with a real-time differentiable renderer, which offers significant speedup to both scene optimization and novel view synthesis.
    Note that for comparable training times to InstantNGP ~\cite{mueller2022instant}, we achieve similar quality to theirs; while this is the maximum quality they reach, by training for 51min we achieve state-of-the-art quality, even slightly better than Mip-NeRF360~\cite{barron2022mipnerf360}. %
    }
	\Description[TeaserFigure]{TeaserFigure}
\end{teaserfigure}
\maketitle

\section{Introduction}

Meshes and points are the most common 3D scene representations because they are explicit and are a good fit for fast GPU/CUDA-based rasterization. 
In contrast, recent Neural Radiance Field (NeRF) methods build on continuous scene representations, typically optimizing a Multi-Layer Perceptron (MLP) using volumetric ray-marching for novel-view synthesis of captured scenes. Similarly, the most efficient radiance field solutions to date build on continuous representations by interpolating values stored in, e.g., voxel~\cite{plenoxels} or hash~\cite{mueller2022instant} grids or points~\cite{xu2022point}.
While the continuous nature of these methods helps optimization, the stochastic sampling required for rendering is costly and can result in noise.
We introduce a new approach that combines the best of both worlds: our 3D Gaussian representation allows optimization with state-of-the-art (SOTA) visual quality and competitive training times, 
while our tile-based splatting solution ensures real-time rendering at SOTA quality for 1080p resolution on several previously published datasets~\cite{knapitsch2017tanks,hedman2018deep,barron2022mipnerf360} (see Fig.~\ref{fig:teaser}).

Our goal is to allow real-time rendering for scenes captured with multiple photos, and create the representations with optimization times as fast as the most efficient previous methods for typical real scenes.
Recent methods achieve fast training~\cite{plenoxels,mueller2022instant}, but struggle to achieve the visual quality obtained by the current SOTA NeRF methods, i.e., Mip-NeRF360~\cite{barron2022mipnerf360}, which requires up to 48 hours of training time. The fast -- but lower-quality -- radiance field methods can achieve interactive rendering times depending on the scene (10-15 frames per second), but fall short of real-time rendering \CORRECTION{($\geq$ 30~fps)}{} at high resolution.

Our solution builds on three main components.  We first introduce \emph{3D Gaussians} as a flexible and expressive scene representation.
We start with the same input as previous NeRF-like methods, i.e., cameras calibrated with Structure-from-Motion (SfM) \cite{snavely2006photo} and initialize the set of 3D Gaussians with the sparse point cloud produced for free as part of the SfM process. In contrast to most point-based solutions that require Multi-View Stereo (MVS) data~\cite{aliev20,kopanas21,ruckert22}, we achieve high-quality results with only SfM points as input. Note that for the NeRF-synthetic dataset, our method achieves high quality even with random initialization.
We show that 3D Gaussians are an excellent choice, since they are a differentiable volumetric representation, but they can also be rasterized very efficiently by projecting them to 2D, and applying standard $\alpha$-blending, using an equivalent image formation model as NeRF.
The second component of our method is optimization of the properties of the 3D Gaussians -- 3D position, opacity $\alpha$, anisotropic covariance, and spherical harmonic (SH) coefficients -- interleaved with adaptive density control steps, where we add and occasionally remove 3D Gaussians during optimization. The optimization procedure produces a reasonably compact, unstructured, and precise representation of the scene (1-5 million Gaussians for all scenes tested).
The third and final element of our method is our real-time rendering solution that uses fast GPU sorting algorithms and is inspired by tile-based rasterization, following recent work~\cite{Lassner_2021_CVPR}. However, thanks to our 3D Gaussian representation, we can perform anisotropic splatting that respects visibility ordering -- thanks to sorting and $\alpha$-blending -- and enable a fast and accurate backward pass by tracking the traversal of as many sorted splats as required.

\noindent
To summarize, we provide the following contributions:
\begin{itemize}
	\item The introduction of anisotropic 3D Gaussians as a high-quality, unstructured representation of radiance fields.
	\item An optimization method of 3D Gaussian properties, interleaved with adaptive density control that creates high-quality representations for captured scenes.
	\item A fast, differentiable rendering approach for the GPU, which is visibility-aware, allows anisotropic splatting and fast backpropagation to achieve high-quality novel view synthesis.
\end{itemize}

\noindent
Our results on previously published datasets show that we can optimize our 3D Gaussians from multi-view captures and achieve equal or better quality than the best quality previous implicit radiance field approaches. We also can achieve training speeds and quality similar to the fastest methods and importantly provide the first \emph{real-time rendering} with high quality for novel-view synthesis.

\section{Related Work}

We first briefly overview traditional reconstruction, then discuss point-based rendering and radiance field work, discussing their similarity; radiance fields are a vast area, so we focus only on directly related work. 
For complete coverage of the field, please see the excellent recent surveys~\cite{tewari2022advances,xie2022neural}.

\subsection{Traditional Scene Reconstruction and Rendering}

The first novel-view synthesis approaches were based on light fields, first densely sampled~\cite{gortler1996lumigraph,levoy1996light} then allowing unstructured capture~\cite{buehler2001unstructured}. The advent of Structure-from-Motion (SfM)~\cite{snavely2006photo} enabled an entire new domain where a collection of photos could be used to synthesize novel views. SfM estimates a sparse point cloud during camera calibration, that was initially used for simple visualization of 3D space. Subsequent multi-view stereo (MVS) produced impressive full 3D reconstruction algorithms over the years ~\cite{goesele2007multi}, enabling the development of several view synthesis algorithms \cite{eisemann2008floating, chaurasia2013depth, hedman2018deep,kopanas21}.
All these methods \emph{re-project} and \emph{blend} the input images into the novel view camera, and use the geometry to guide this re-projection. These methods produced excellent results in many cases, but typically cannot completely recover from unreconstructed regions, or from ``over-reconstruction'', when MVS generates inexistent geometry.
Recent neural rendering algorithms ~\cite{tewari2022advances} vastly reduce such artifacts and avoid the overwhelming cost of storing all input images on the GPU, outperforming these methods on most fronts.

\subsection{Neural Rendering and Radiance Fields}

Deep learning techniques were adopted early for novel-view synthesis ~\cite{zhou2016view, flynn2016deepstereo}; CNNs were used to estimate blending weights~\cite{hedman2018deep}, or for texture-space solutions \cite{thies2019deferred,riegler2020free}. The use of MVS-based geometry is a major drawback of most of these methods; in addition, the use of CNNs for final rendering \ADDITION{frequently} results in \REMOVAL{frequent} temporal flickering.

Volumetric representations for novel-view synthesis were initiated by Soft3D~\cite{penner2017soft}; deep-learning techniques coupled with volumetric ray-marching were subsequently proposed ~\cite{sitzmann2019deepvoxels,henzler2019escaping} building on a continuous differentiable density field to represent geometry. Rendering using volumetric ray-marching has a significant cost due to the large number of samples required to query the volume. Neural Radiance Fields (NeRFs)~\cite{mildenhall2020nerf} introduced importance sampling and positional encoding to improve quality, but used a large Multi-Layer Perceptron negatively affecting speed.
The success of NeRF has resulted in an explosion of follow-up methods that address quality and speed, often by introducing regularization strategies; the current state-of-the-art in image quality for novel-view synthesis is Mip-NeRF360~\cite{barron2022mipnerf360}. While the rendering quality is outstanding, training and rendering times remain extremely high; we are able to equal or in some cases surpass this quality while providing fast training and real-time rendering.

The most recent methods have focused on faster training and/or rendering mostly by exploiting three design choices: the use of spatial data structures to store (neural) features that are subsequently interpolated during volumetric ray-marching, different encodings, and MLP capacity. Such methods include
different variants of space discretization~\cite{nglod-cvpr2021,plenoxels,yu2021plenoctrees,hedman2021snerg,chen2022mobilenerf,garbin2021fastnerf,Reiser2021ICCV,tensorf-eccv2022,wu2022snisr}, 
codebooks~\cite{takikawa2022variable}, 
and encodings such as hash tables~\cite{mueller2022instant}, allowing the use of a smaller MLP or foregoing neural networks completely \cite{plenoxels,dvgo-cvpr2022}. %

Most notable of these methods are InstantNGP~\cite{mueller2022instant} which uses a hash grid and an occupancy grid to accelerate computation and a smaller MLP to represent density and appearance;
and Plenoxels~\cite{plenoxels} that use a sparse voxel grid to interpolate a continuous density field, and are able to forgo neural networks altogether. \CORRECTION{They both replace the MLP used to represent directional effects with much faster Spherical Harmonics.}
{Both rely on Spherical Harmonics: the former to represent directional effects directly, the latter to encode its inputs to the color network.}
While both provide outstanding results, these methods can still struggle to represent empty space effectively, depending in part on the scene/capture type.  In addition, image quality is limited in large part by the choice of the structured grids used for acceleration, and rendering speed is hindered by the need to query many samples for a given ray-marching step. The unstructured, explicit GPU-friendly 3D Gaussians we use %
achieve faster rendering speed and %
better quality \emph{without} neural components.

\subsection{Point-Based Rendering and Radiance Fields}

Point-based methods efficiently render disconnected and unstructured geometry samples (i.e., point clouds)~\cite{gross2011point}. 
In its simplest form, point sample rendering \cite{Grossman1998PointSR} rasterizes an unstructured set of points with a fixed size, for which it may exploit natively supported point types of graphics APIs \cite{SAINZ2004869} or parallel software rasterization on the GPU \cite{10.1145/3543863, laine2011high}. While true to the underlying data, point sample rendering suffers from holes, causes aliasing, and is strictly discontinuous. Seminal work on high-quality point-based rendering addresses these issues by “splatting” point primitives with an extent larger than a pixel, e.g., circular or elliptic discs, ellipsoids, or surfels \cite{10.1145/383259.383300, 10.5555/2386366.2386369,  Ren2002ObjectSE, 10.1145/344779.344936}.

There has been recent interest in \emph{differentiable} point-based rendering techniques~\cite{yifan19,wiles2020synsin}. Points have been augmented with neural features and rendered using a CNN~\cite{ruckert22,aliev20} resulting in fast or even real-time view synthesis; however they still depend on MVS for the initial geometry, and as such inherit its artifacts, most notably over- or under-reconstruction in hard cases such \CORRECTION{}{as} featureless/shiny areas or thin structures. 

Point-based $\alpha$-blending and NeRF-style volumetric rendering share essentially the same image formation model.
Specifically, the color $C$ is given by volumetric rendering along a ray:
\begin{equation}
	\label{eq:nerf-rendering-quadrature}
	C = \sum_{i=1}^N T_i(1-\exp(-\sigma_i\delta_i))\mathbf{c}_i \hspace{0.5em} \text{ with } \hspace{0.5em} T_i = \exp\left(-\sum_{j=1}^{i-1}\sigma_j\delta_j\right),
\end{equation}
where samples of density $\sigma$, transmittance $T$, and color $\mathbf{c}$ are taken along the ray with intervals $\delta_i$.
This can be re-written as
\begin{equation}
	\label{eq:nerf-rendering-quadrature2}
	C = \sum_{i=1}^N T_i\alpha_i\mathbf{c}_i,
\end{equation}
with
\begin{equation*}
	\alpha_i = (1-\exp(-\sigma_i\delta_i)) 
	\hspace{0.5em} \text{and} \hspace{0.5em}
	T_i = \prod_{j=1}^{i-1}(1-\alpha_i).
\end{equation*}
A typical neural point-based approach (e.g.,~\cite{kopanas21,kopanas22}) computes the color $C$ of a pixel by blending $\mathcal{N}$ ordered points overlapping the pixel:
\begin{equation}
	\label{eq:front-to-back}
	C = \sum_{i \in \mathcal{N}}
	c_{i}\alpha_{i}
	\prod_{j=1}^{i-1}(1-\alpha_{j}),
\end{equation}
where $\mathbf{c}_i$ is the color of each point and $\alpha_i$ is given by evaluating a 2D Gaussian with covariance $\Sigma$ ~\cite{yifan19} multiplied with a learned per-point opacity. 

From Eq.~\ref{eq:nerf-rendering-quadrature2} and Eq.~\ref{eq:front-to-back}, we can clearly see that the image formation model is the same. However, the rendering algorithm is very different. NeRFs are a continuous representation implicitly representing empty/occupied space; expensive random sampling is required to find the samples in Eq.~\ref{eq:nerf-rendering-quadrature2} with consequent noise and computational expense. In contrast, points are an unstructured, discrete representation that is flexible enough to allow creation, destruction, and displacement of geometry similar to NeRF. This is achieved by optimizing opacity and positions, as shown by previous work~\cite{kopanas21}, while avoiding the shortcomings of a full volumetric representation. 

Pulsar~\cite{Lassner_2021_CVPR} achieves fast \emph{sphere} rasterization which inspired our tile-based and sorting renderer. However, given the analysis above, we want to maintain (approximate) conventional $\alpha$-blending on sorted splats to have the advantages of volumetric representations: Our rasterization respects visibility order in contrast to their order-independent method. In addition, 
\CORRECTION{}{we} backpropagate gradients on all splats in a pixel and rasterize anisotropic splats.
These elements all contribute to the high visual quality of our results (see Sec.~\ref{sec:ablations}).
In addition, 
previous methods mentioned above
also use CNNs for rendering, which results in temporal instability.
\CORRECTION{}{Nonetheless, the rendering speed of Pulsar~\cite{Lassner_2021_CVPR} and ADOP~\cite{ruckert22} served as motivation to develop our fast rendering solution.}

While focusing on specular effects, the diffuse point-based rendering track of Neural Point Catacaustics~\cite{kopanas22} overcomes this temporal instability by using an MLP, but still required MVS geometry as input.
The most recent method~\cite{zhang2022} in this category does not require MVS, and also uses SH for directions; however, it can only handle scenes of one object and needs masks for initialization. While fast for small resolutions and low point counts, it is unclear how it can scale to scenes of typical datasets~\cite{knapitsch2017tanks,hedman2018deep,barron2022mipnerf360}.
We use 3D Gaussians for a more flexible scene representation, avoiding the need for MVS geometry and achieving real-time rendering thanks to our tile-based rendering algorithm for the projected Gaussians.

A recent approach~\cite{xu2022point} uses points to represent a radiance field with a radial basis function approach. They employ point pruning and densification techniques during optimization, but use volumetric ray-marching and cannot achieve real-time display rates. %

In the domain of human performance capture, 3D Gaussians have been used to represent captured human bodies~\cite{stoll2011fast,rhodin2015versatile}; more recently \ADDITION{they have been used with volumetric ray-marching for vision tasks~\cite{voge}.} \CORRECTION{neural}{Neural} volumetric primitives have been proposed in a similar context~\cite{lombardi2021mixture}. \CORRECTION{While we also use}{While these methods inspired the choice of} 3D Gaussians as our scene representation, \CORRECTION{these methods}{they} focus on the specific case of reconstructing and rendering a single isolated object (a human body or face), resulting in scenes with small depth complexity. In contrast, our optimization of \emph{anisotropic} covariance, our interleaved optimization/density control, and efficient depth sorting for rendering allow us to handle complete, complex scenes including background, both indoors and outdoors and with large depth complexity.

\begin{figure*}[!h]
	\includegraphics[width=\linewidth]{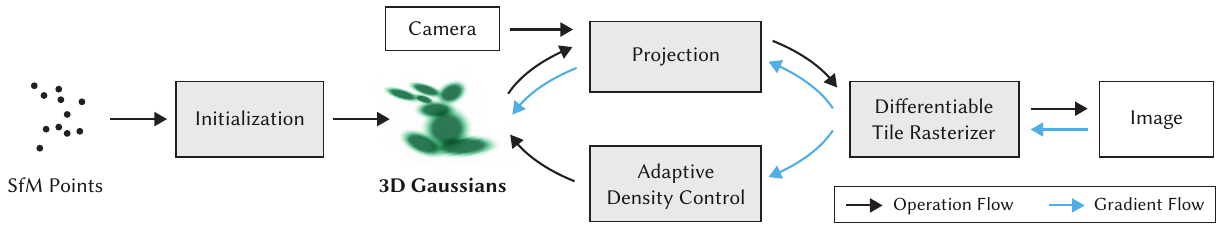}
	\caption{
		Optimization starts with the sparse SfM point cloud and creates a set of 3D Gaussians. We then optimize and adaptively control the density of this set of Gaussians. During optimization we use our fast tile-based renderer, allowing competitive training times compared to SOTA fast radiance field methods. Once trained, our renderer allows real-time navigation for a wide variety of scenes.
	}
	\label{fig:overview}
\end{figure*}
\section{Overview}
The input to our method is a set of images of a static scene, together with the corresponding cameras calibrated by SfM \cite{schoenberger2016sfm} which produces a sparse point cloud as a side-effect.
From these points we create a set of 3D Gaussians %
(Sec.~\ref{sec:3d-splats}), 
defined by a position (mean), covariance matrix and opacity $\alpha$, that allows a very flexible optimization regime. This results in a reasonably compact representation of the 3D scene, in part because highly anisotropic volumetric splats can be used to represent fine structures compactly. The directional appearance component (color) of the radiance field is represented via spherical harmonics (SH), following standard practice~\cite{plenoxels,mueller2022instant}. Our algorithm proceeds to create the radiance field representation (Sec.~\ref{sec:opt-dens}) via a sequence of optimization steps of 3D Gaussian parameters, i.e., position, covariance, $\alpha$ and SH coefficients interleaved with operations for adaptive control of the Gaussian density.
The key to the efficiency of our method is our tile-based rasterizer (Sec.~\ref{sec:tile-raster}) that allows $\alpha$-blending of anisotropic splats, respecting visibility order thanks to fast sorting. Out fast rasterizer also includes a fast backward pass by tracking accumulated $\alpha$ values, without a limit on the number of Gaussians that can receive gradients. 
The overview of our method is illustrated in Fig.~\ref{fig:overview}.

\section{Differentiable 3D Gaussian Splatting}
\label{sec:3d-splats}

Our goal is to optimize a scene representation that allows high-quality novel view synthesis, starting from a sparse set of (SfM) points without normals. To do this, we need a primitive that inherits the properties of differentiable volumetric representations, while at the same time being unstructured and explicit to allow very fast rendering. We choose 3D Gaussians, which are differentiable
and can be easily projected to 2D splats allowing fast $\alpha$-blending for rendering.

Our representation has similarities to
previous methods that use 2D points ~\cite{yifan19,kopanas21} and assume each point is a small planar circle with a normal. 
Given the extreme sparsity of SfM points it is very hard to estimate normals. Similarly, optimizing very noisy normals from such an estimation would be very challenging.
Instead, we model the geometry as a set of 3D Gaussians that do not require normals. Our Gaussians are defined by a full 3D covariance matrix $\Sigma$ defined in world space~\cite{zwicker2001ewa} centered at point (mean) $\mu$:
\begin{equation}
	G(x)~= e^{-\frac{1}{2}(x)^{T}\Sigma^{-1}(x)}
\end{equation}
\CORRECTION{and $|\Sigma|$ is the determinant of $\Sigma$ that is a symmetric 3$\times$3 matrix}{}. This Gaussian is multiplied by $\alpha$ in our blending process.

However, we need to project our 3D Gaussians to 2D for rendering.
Zwicker et al. ~\shortcite{zwicker2001ewa} demonstrate how to do this projection to image space. Given a viewing transformation $W$
the covariance matrix $\Sigma'$ in camera coordinates is given as follows:
\begin{equation}
	\label{eq:volume-render}
	\Sigma' = J W ~\Sigma ~W ^{T}J^{T}
\end{equation}
where $J$ is the Jacobian of the affine approximation of the projective transformation. \CORRECTION{The authors}{Zwicker et al. ~\shortcite{zwicker2001ewa}} also show that if we skip the third row and column of $\Sigma'$, 
	we obtain a 2$\times$2 variance matrix with the same structure and properties as if we would start from planar points with normals, as in previous work~\cite{kopanas21}.
	
	An obvious approach would be to directly optimize the covariance matrix $\Sigma$ to obtain 3D Gaussians that represent the radiance field. However, covariance matrices have physical meaning only when they are positive semi-definite. 
	For our optimization of all our parameters, we use gradient descent that cannot be easily constrained to produce such valid matrices, and update steps and gradients can very easily create invalid covariance matrices. 
	
	As a result, we opted for a more intuitive, yet equivalently expressive representation for optimization.
	The covariance matrix $\Sigma$ of a 3D Gaussian is analogous to describing the configuration of an ellipsoid.
	Given a scaling matrix $S$ and rotation matrix $R$, we can find the corresponding $\Sigma$:
	\begin{equation}
		\Sigma = RSS^TR^T
	\end{equation}
	
	To allow independent optimization of both factors, we store them separately: a 3D vector $s$ for scaling and a quaternion $q$ to represent rotation. These can be trivially converted to their respective matrices and combined, making sure to normalize $q$ to obtain a valid unit quaternion.
	
	To avoid significant overhead due to automatic differentiation during training, we derive the gradients for all parameters explicitly. Details of the exact derivative computations are in appendix~\ref{sec:appa}.
	\CORRECTION{We discuss the specific case of scaling and rotation (i.e., the shape) of 3D Gaussians next.}{}
	\CORRECTION{The differentiable renderer outputs 
	the $2\times2$ screen-space covariance matrix $\Sigma'$.
	Given the gradient of the loss $\frac{dL}{d\Sigma'}$ %
	which we assume as output from the differentiable renderer, 
	we can apply the chain rule to compute gradients and propagate the loss.}{}

\begin{figure}[!h]
		\begin{overpic}[width=\columnwidth]{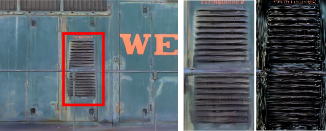}
			\put (61,3) {\color{white}Original}
			\put (82.2,5) {\color{white}Shrunken} 
			\put (82,1.2) {\color{white}Gaussians}
		\end{overpic}
		
		\caption{
			We visualize the 3D Gaussians after optimization by shrinking them 60\% (far right). This clearly shows the anisotropic shapes of the 3D Gaussians that compactly represent complex geometry after optimization. Left the actual rendered image.
		}
		\label{fig:aniso-cov}
	\end{figure}

	This representation of anisotropic covariance -- suitable for optimization -- allows us to optimize 3D Gaussians to adapt to the geometry of different shapes in captured scenes, resulting in a fairly compact representation. 
	Fig.~\ref{fig:aniso-cov} illustrates such cases.

\section{Optimization with Adaptive Density Control of 3D Gaussians}
\label{sec:opt-dens}

The core of our approach is the optimization step, which creates a dense set of 3D Gaussians accurately representing the scene for free-view synthesis.
In addition to positions $p$, $\alpha$, and covariance $\Sigma$, we also optimize SH coefficients representing color $c$ of each Gaussian to correctly capture the view-dependent appearance of the scene. 
The optimization of these parameters is interleaved with steps that control the density of the Gaussians to better represent the scene.

\subsection{Optimization}

The optimization is based on successive iterations of rendering and comparing the resulting image to the training views in the captured dataset. Inevitably, geometry may be incorrectly placed due to the ambiguities of 3D to 2D projection. Our optimization thus needs to be able to \emph{create} geometry and also \emph{destroy} or \emph{move} geometry if it has been incorrectly positioned. The quality of the parameters of the covariances of the 3D Gaussians is critical for the compactness of the representation since large homogeneous areas can be captured with a small number of large anisotropic Gaussians.

We use Stochastic Gradient Descent techniques for optimization, taking full advantage of standard GPU-accelerated frameworks\TODO{ADD pytorch ref}, and the ability to add custom CUDA kernels for some operations, following recent best practice~\cite{plenoxels,dvgo-cvpr2022}. In particular, our fast rasterization (see Sec.~\ref{sec:tile-raster}) is critical in the efficiency of our optimization, since it is the main computational bottleneck of the optimization.

We use a sigmoid activation function for $\alpha$ to constrain it in the $[0-1)$ range and obtain smooth gradients, and an exponential activation function for the scale of the covariance for similar reasons.

We estimate the initial covariance matrix as an isotropic Gaussian with axes
equal to the mean of the distance to the closest three points. 
We use a standard exponential decay scheduling technique similar to Plenoxels~\cite{plenoxels}, but for positions only. The loss function is $\mathcal{L}_1$ combined with a D-SSIM term:

\begin{equation}
	\mathcal{L} = (1 - \lambda) \mathcal{L}_1 + \lambda \mathcal{L_{\textrm{D-SSIM}}}
\end{equation}

We use $\lambda~=~0.2$ in all our tests.
We provide details of the learning schedule and other elements in Sec.~\ref{sec:impl}.

\subsection{Adaptive Control of Gaussians}
We start with the initial set of sparse points from SfM and then apply 
our method to adaptively control the \REMOVAL{density \Dg~}\CORRECTION{of the 3D Gaussians}{number of Gaussians and their density over unit volume}\footnote{Density \CORRECTION{\Dg~}{of Gaussians} should not be confused of course with density $\sigma$ in the NeRF literature.}, allowing us to go from an initial sparse set of Gaussians to a denser set that better represents the \CORRECTION{screen}{scene}, and with correct parameters. 
After optimization warm-up (see Sec.~\ref{sec:impl}), we densify every 100 iterations and remove any Gaussians that are essentially transparent, i.e., with $\alpha$ less than a threshold $\epsilon_{\alpha}$.

Our adaptive control of \CORRECTION{density \Dg~}{the Gaussians} needs to populate empty areas. It focuses on regions with missing geometric features (``under-reconstruction''), but also in regions where Gaussians cover large areas in the scene (which often correspond to ``over-reconstruction'').
\TODO{MAYBE ADD FIGURE (GK)}
We observe that both have \emph{large} view-space positional gradients.
Intuitively, this is likely because they correspond to regions that are not yet well reconstructed, and the optimization tries to move the Gaussians to correct this. 

Since both cases are good candidates for densification, \CORRECTION{so}{} we 
densify Gaussians with an average magnitude of view-space position gradients above a threshold~$\tau_{\textrm{pos}}$, which we set to $0.0002$ in our tests.

\begin{figure}[!h]
	\includegraphics[width=\linewidth]{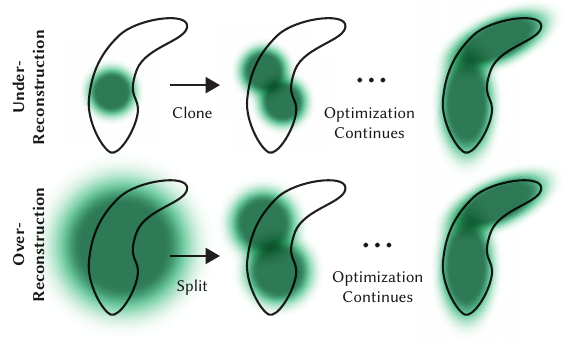}
	\caption{
		\label{fig:density_control}
		Our adaptive Gaussian densification scheme.
		\emph{Top row (under-reconstruction)}: When small-scale geometry (black outline) is insufficiently covered, we clone the respective Gaussian.
		\emph{Bottom row (over-reconstruction)}: If small-scale geometry is represented by one large splat, we split it in two.
	}
\end{figure}

We next present details of this process, illustrated in Fig.~\ref{fig:density_control}.

For small Gaussians that are in under-reconstructed regions, we need to cover the new geometry that must be created. 
For this, it is preferable to clone the Gaussians, by simply creating a copy of the same size, and moving it in the direction of the positional gradient.

On the other hand, large Gaussians in regions with high variance need to be split into smaller Gaussians. We replace such Gaussians by\REMOVAL{thus delete such Gaussians and introduce} two new ones\REMOVAL{Gaussians}, and divide their scale by a factor of $\phi~=~1.6$ which we determined experimentally.
We also initialize their position by 
using the original 3D Gaussian as a PDF for sampling.

In the first case we detect and treat the need for \CORRECTION{increasing the density \Dg}{increasing both the total volume of the system and the number of Gaussians}, while in the second case we conserve \CORRECTION{the total density \Dg~ of the system.}{total volume but increase the number of Gaussians.} 
\CORRECTION{Given that our problem is not strictly convex, the optimization sometimes can get stuck in local minima which may result in an unjustified increase in the Gaussian density.}{Similar to other volumetric representations, our optimization can get stuck with floaters close to the input cameras; in our case this may result in an unjustified increase in the Gaussian density.}
An effective way to moderate the increase in the number of Gaussians \CORRECTION{and to deal with the floaters} is to set the $\alpha$ value close to zero every $N=3000$ iterations. The optimization then increases the $\alpha$ for the Gaussians where this is needed while allowing our culling approach to remove Gaussians with $\alpha$ less than $\epsilon_{\alpha}$ as described above. \CORRECTION{We also}{Gaussians may shrink or grow and considerably overlap with others, but we periodically} remove \CORRECTION{}{Gaussians that are} very large \CORRECTION{Gaussians} in worldspace and \CORRECTION{Gaussians}{those} that have a big footprint in viewspace. This strategy results in overall good control over the total number of Gaussians. \ADDITION{The Gaussians in our model remain primitives in Euclidean space at all times; unlike other methods~\cite{barron2022mipnerf360,plenoxels}, we do not require space compaction, warping or projection strategies for distant or large Gaussians.}

\section{Fast Differentiable Rasterizer for Gaussians}
\label{sec:tile-raster}

Our goals are to have fast overall rendering and fast sorting to allow approximate $\alpha$-blending -- including for anisotropic splats -- and to avoid hard limits on the number of splats that can receive gradients that exist in previous work~\cite{Lassner_2021_CVPR}.

To achieve these goals, we design a tile-based rasterizer for Gaussian splats inspired by recent software rasterization approaches~\cite{Lassner_2021_CVPR} to pre-sort primitives for an entire image at a time, avoiding the expense of sorting per pixel that hindered previous $\alpha$-blending solutions~\cite{kopanas21,kopanas22}. Our fast rasterizer allows efficient backpropagation over an arbitrary number of blended Gaussians with low additional memory consumption, requiring only a constant overhead per pixel. 
Our rasterization pipeline is fully differentiable, and given the projection to 2D (Sec.~\ref{sec:3d-splats}) can rasterize anisotropic splats similar to previous 2D splatting methods~\cite{kopanas21}.

Our method starts by splitting the screen into 16$\times$16 tiles, and then proceeds to cull 3D Gaussians against the view frustum and each tile. 
Specifically, we only keep Gaussians with a 99\% confidence interval intersecting the view frustum. 
Additionally, we use a guard band to trivially reject Gaussians at extreme positions \ADDITION{(i.e., those with means close to the near plane and far outside the view frustum)}, since computing their projected 2D covariance would be unstable.
We then instantiate each Gaussian according to the number of tiles they overlap and assign each instance a key that combines view space depth and tile ID. We then sort Gaussians based on these keys using a single fast GPU Radix sort~\cite{merrill2010revisiting}.
Note that there is no additional per-pixel ordering of points, and blending is performed based on this initial sorting.  As a consequence, our $\alpha$-blending can be approximate in some configurations. %
However, these approximations become negligible as \CORRECTION{points}{splats} approach the size of individual pixels. 
We found that this choice greatly enhances training and rendering performance without producing visible artifacts in converged scenes. 

After sorting \CORRECTION{points}{Gaussians}, we produce a list for each tile by identifying the first and last depth-sorted entry that splats to a given tile. For rasterization, we launch one thread block for each tile. Each block first collaboratively loads packets of \CORRECTION{points}{Gaussians} into shared memory and then, for a given pixel, accumulates color and $\alpha$ values by traversing the lists front-to-back, thus maximizing the gain in parallelism both for data loading/sharing and processing.
When we reach a target saturation of $\alpha$ in a pixel, the corresponding thread stops. At regular intervals, threads in a tile are queried and the processing of the entire tile terminates when all pixels have saturated (i.e., $\alpha$ goes to 1). %
Details of sorting and a high-level overview of the overall rasterization approach are given in Appendix~\ref{app:raster}.

During rasterization, the saturation of $\alpha$ is the only stopping criterion. In contrast to previous work, we do not limit the number of blended primitives that receive gradient updates. We enforce this property to allow our approach to handle scenes with an arbitrary, varying depth complexity and accurately learn them, without having to resort to scene-specific hyperparameter tuning. 
During the backward pass, we must therefore recover the full sequence of blended points per-pixel in the forward pass. One solution would be to store arbitrarily long lists of blended points per-pixel in global memory~\cite{kopanas21}. To avoid the implied dynamic memory management overhead, we instead choose to traverse the per-tile lists again; we can reuse the sorted array of Gaussians and tile ranges from the forward pass. To facilitate gradient computation, we now traverse them back-to-front.

The traversal starts from the last point that affected any pixel in the tile, and loading of points into shared memory again happens collaboratively. Additionally, each pixel will only start (expensive) overlap testing and processing of points if their \CORRECTION{index}{depth} is lower than or equal to the \CORRECTION{index}{depth} of the last point that contributed to its color during the forward pass.
Computation of the gradients described in Sec.~\ref{sec:3d-splats} requires the accumulated opacity values at each step during the original blending process.
\CORRECTION{ Rather than storing a list of of progressively shrinking opacity values during the forward pass, we recover the values from those in the forward pass. }{Rather than trasversing an explicit list of progressively shrinking opacities in the backward pass, we can recover these intermediate opacities by storing only the total accumulated opacity at the end of the forward pass.}
Specifically, each point stores the final accumulated opacity $\alpha$ in the forward process; we divide this by each point's $\alpha$ in our back-to-front traversal to obtain the required coefficients for gradient computation.

\newcommand{\zoomin}[9]{ %
\begin{tikzpicture}[spy using outlines={rectangle,#9,magnification=#8,size=#6}]   
	\node[anchor=south west,inner sep=0]  {\includegraphics[width=#7]{#1}};
	\spy on (#2, #3) in node at (#4,#5);
\end{tikzpicture}
}

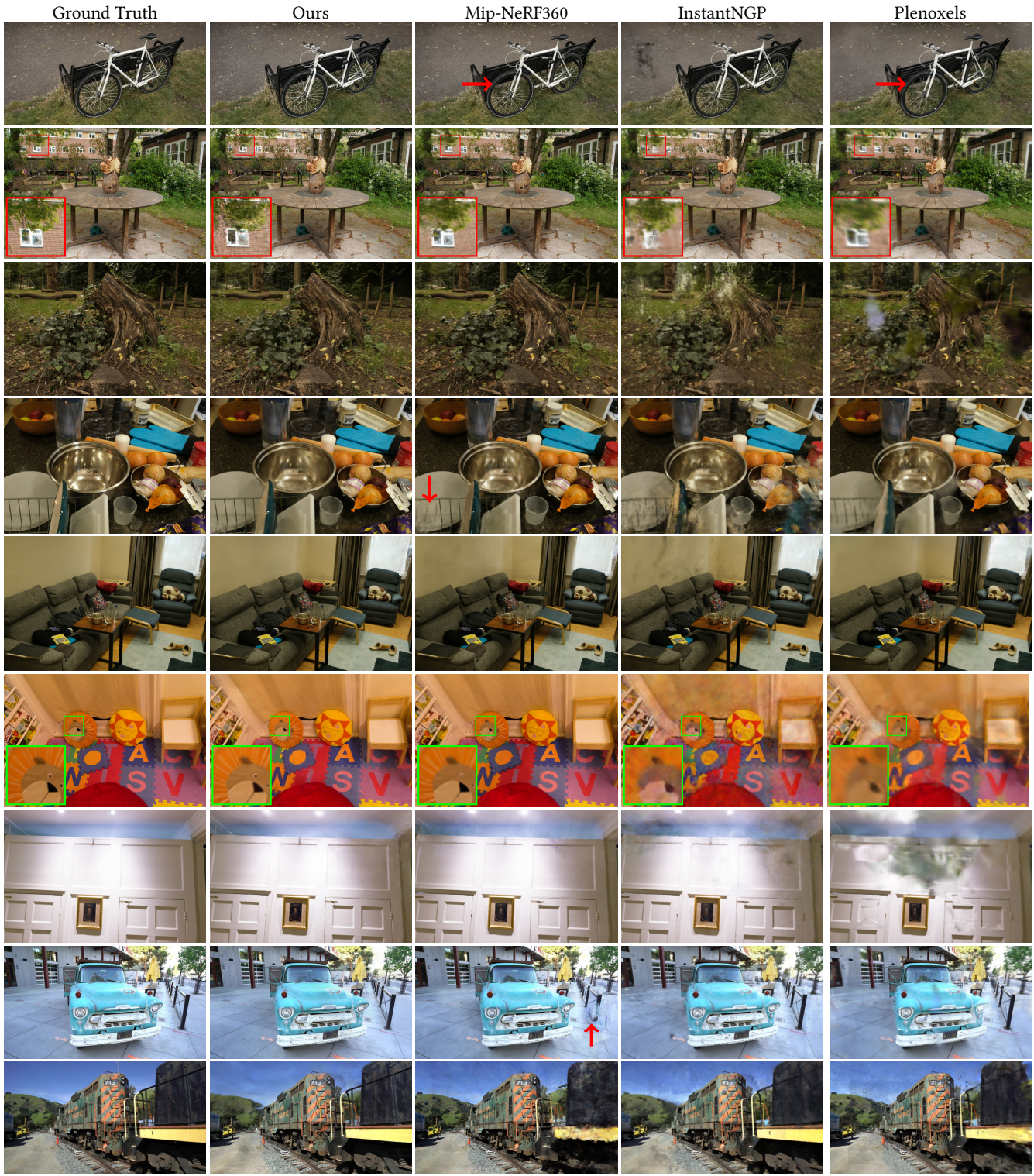
\begin{figure*}[!h]
	\newlength\mytmplen
	\setlength\mytmplen{.193\linewidth}
	\setlength{\tabcolsep}{1pt}
	\renewcommand{\arraystretch}{0.5}
	\centering
	\begin{tabular}{cccccc}
		Ground Truth&Ours& Mip-NeRF360 & InstantNGP & Plenoxels \\
		\includegraphics[width=\mytmplen,trim={0 1.5cm 0 2cm},clip]{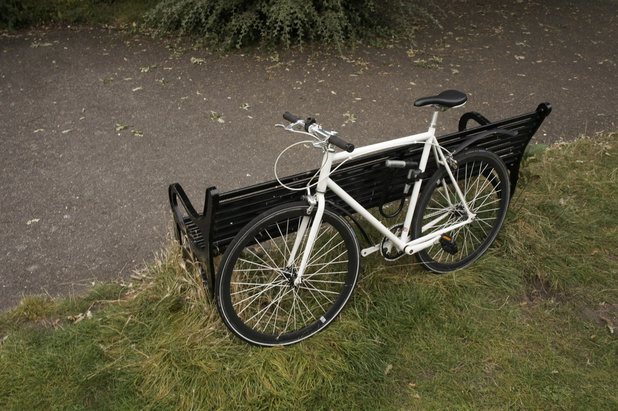} &
		\includegraphics[width=\mytmplen,trim={0 1.5cm 0 2cm},clip]{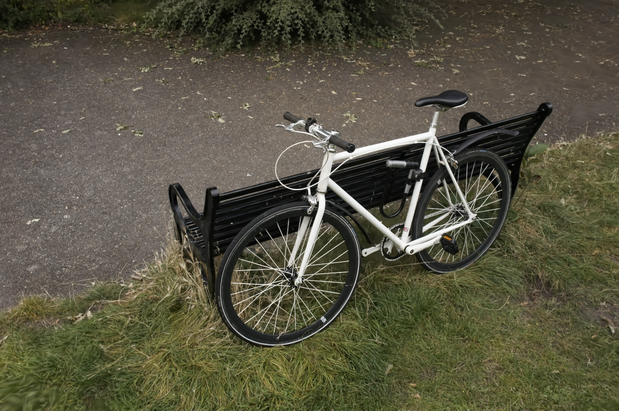} &
	    \begin{tikzpicture}
			\node[anchor=south west,inner sep=0] (image) at (0,0) {\includegraphics[width=\mytmplen,trim={0 1.5cm 0 2cm},clip]{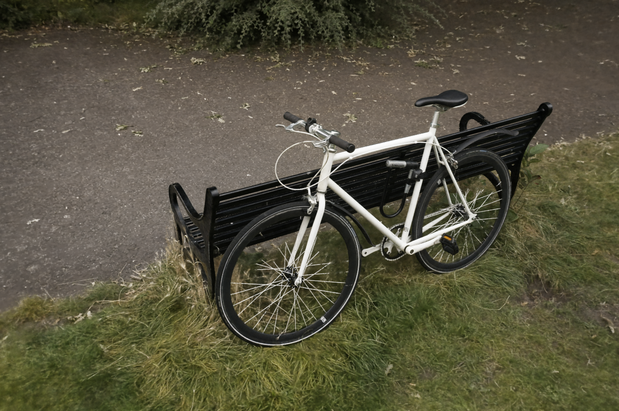}};
			\begin{scope}[x={(image.south east)},y={(image.north west)}]
				\draw[red,ultra thick,->] (0.23,0.40) -- (0.38,0.40);
			\end{scope}
		\end{tikzpicture} &
		\includegraphics[width=\mytmplen,trim={0 1.5cm 0 2cm},clip]{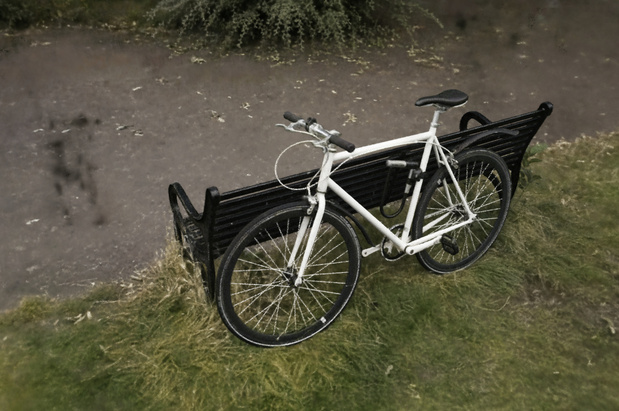} &
			    \begin{tikzpicture}
			\node[anchor=south west,inner sep=0] (image) at (0,0) {\includegraphics[width=\mytmplen,trim={0 1.5cm 0 2cm},clip]{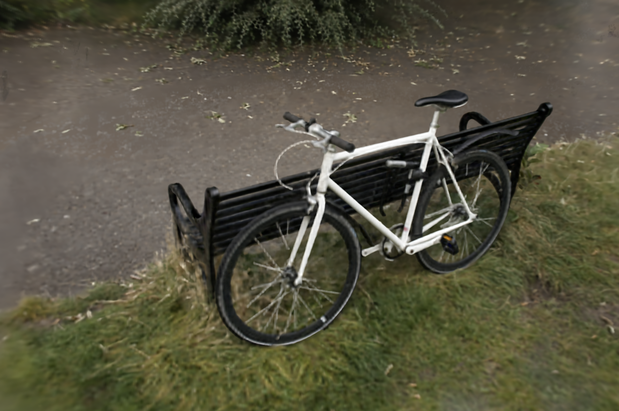}};
			\begin{scope}[x={(image.south east)},y={(image.north west)}]
				\draw[red,ultra thick,->] (0.23,0.40) -- (0.38,0.40);
			\end{scope}
		\end{tikzpicture} \\
		\zoomin{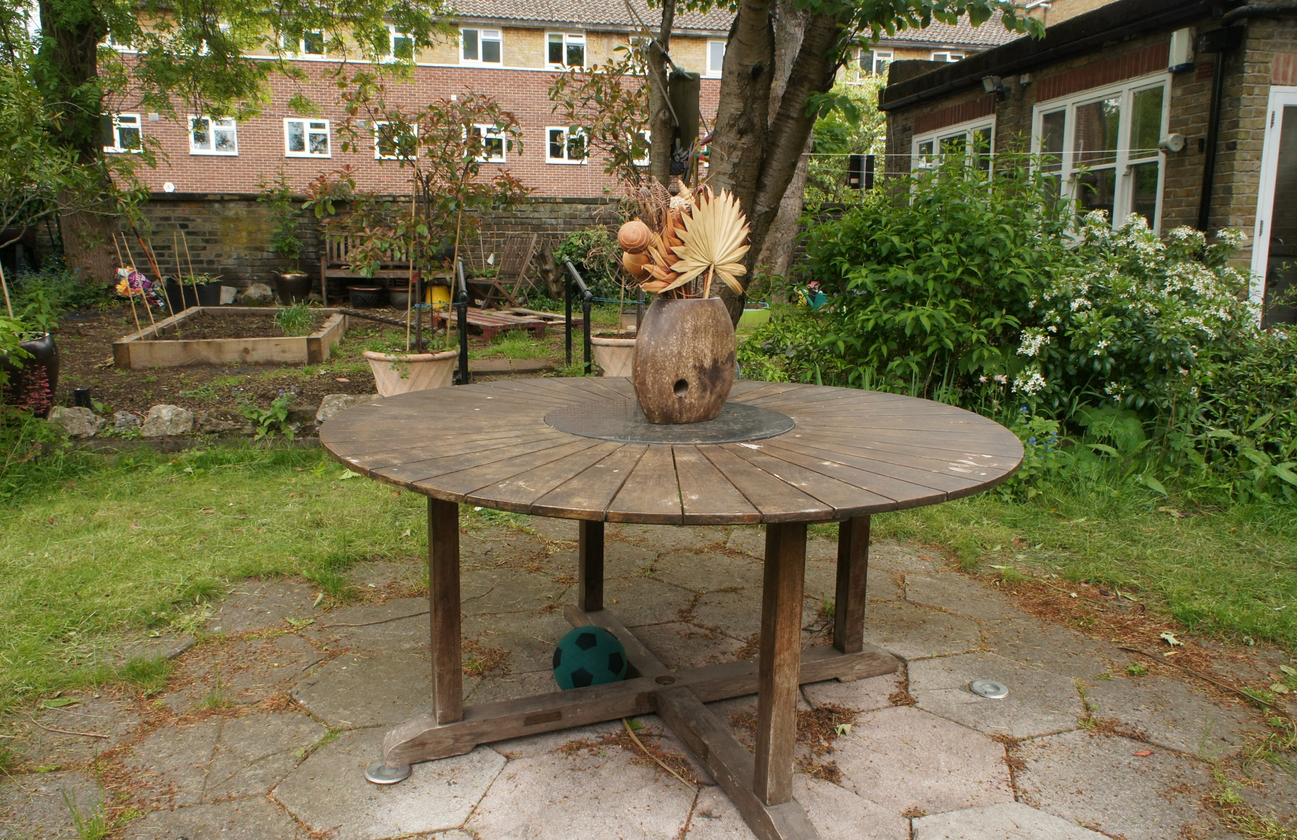}{0.6}{1.95}{0.55cm}{0.55cm}{1cm}{\mytmplen}{3}{red}&
		\zoomin{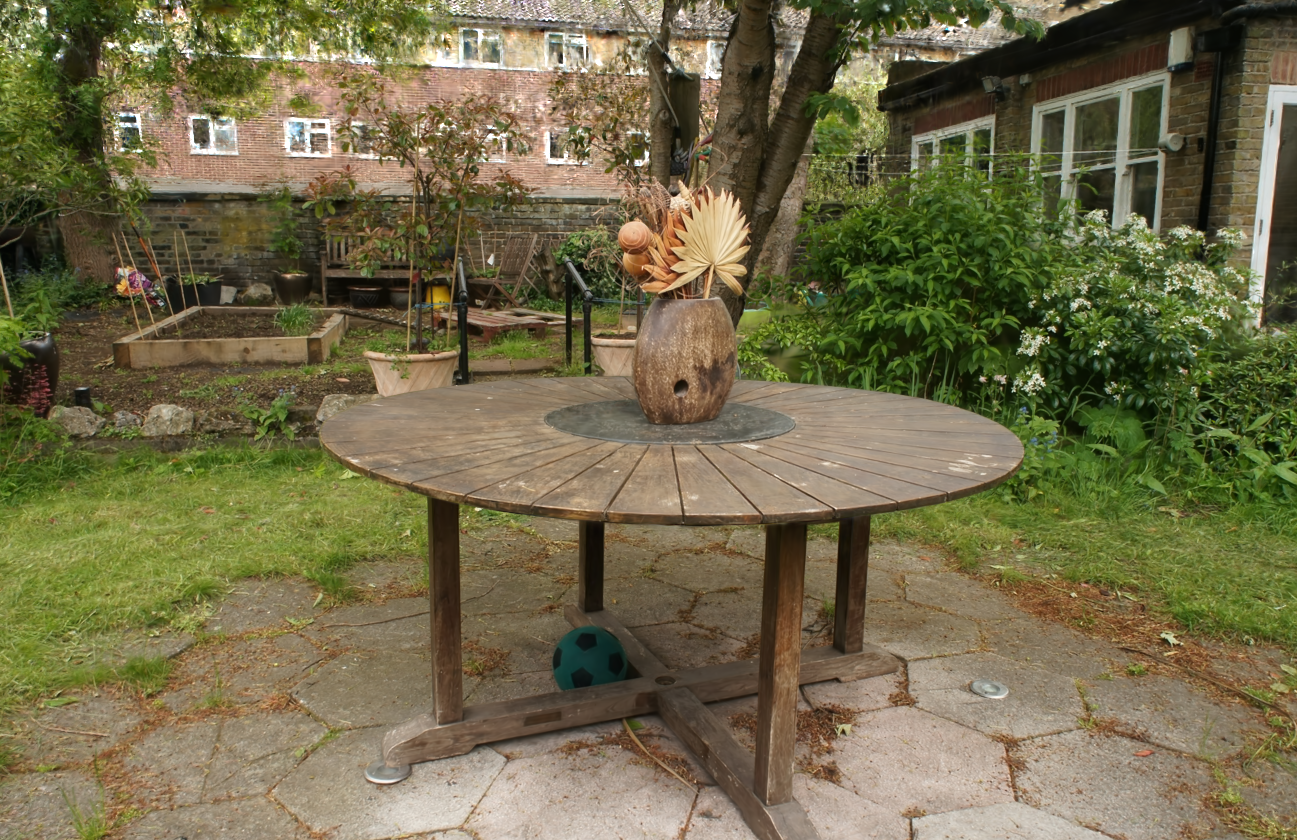}{0.6}{1.95}{0.55cm}{0.55cm}{1cm}{\mytmplen}{3}{red}&
		\zoomin{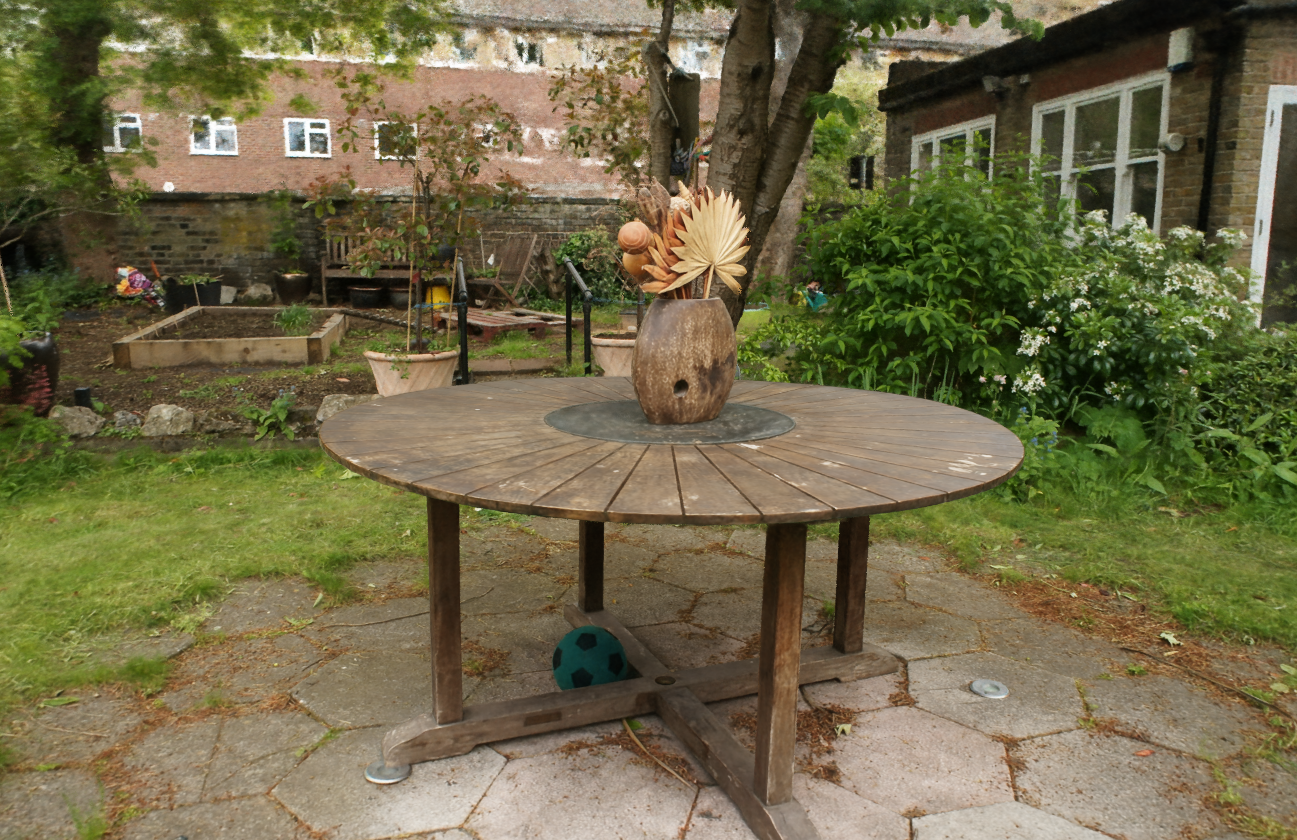}{0.6}{1.95}{0.55cm}{0.55cm}{1cm}{\mytmplen}{3}{red}&
		\zoomin{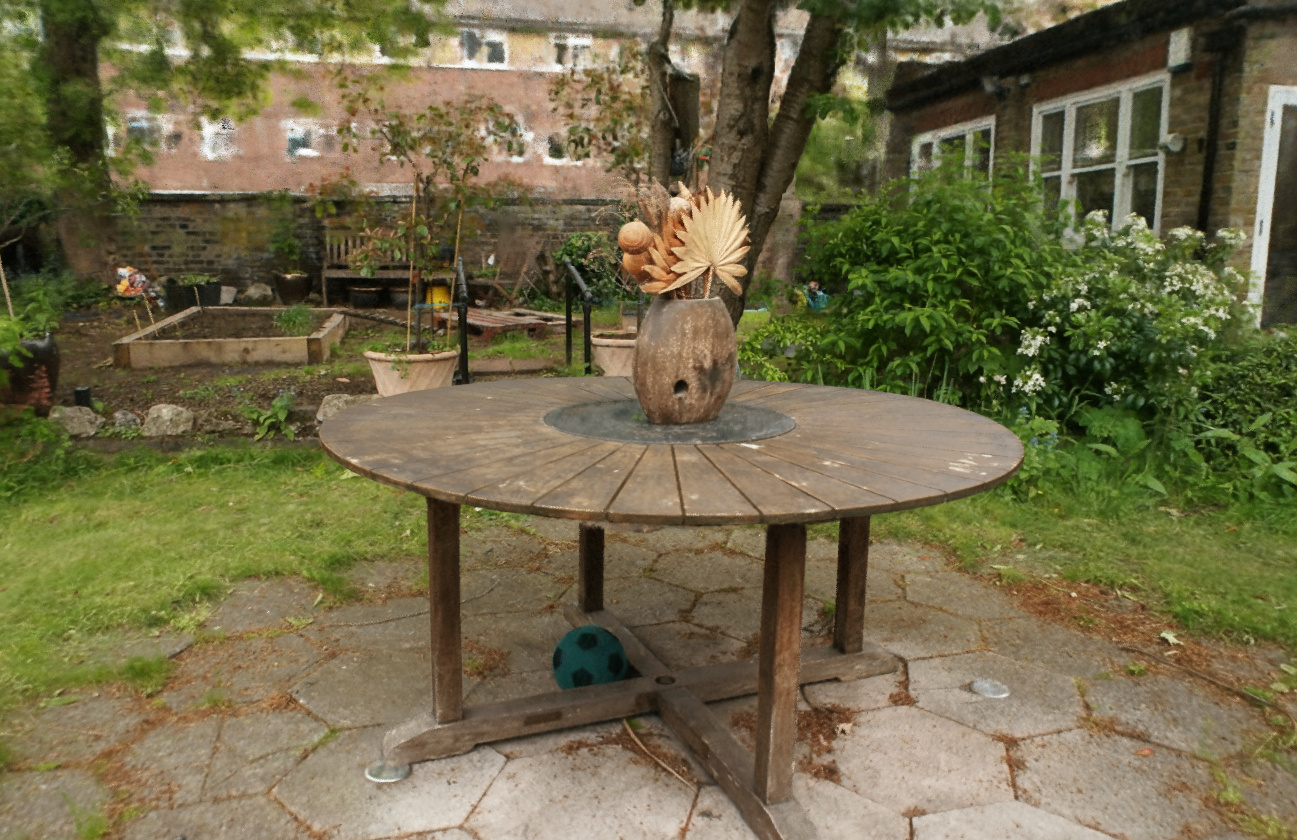}{0.6}{1.95}{0.55cm}{0.55cm}{1cm}{\mytmplen}{3}{red}&
		\zoomin{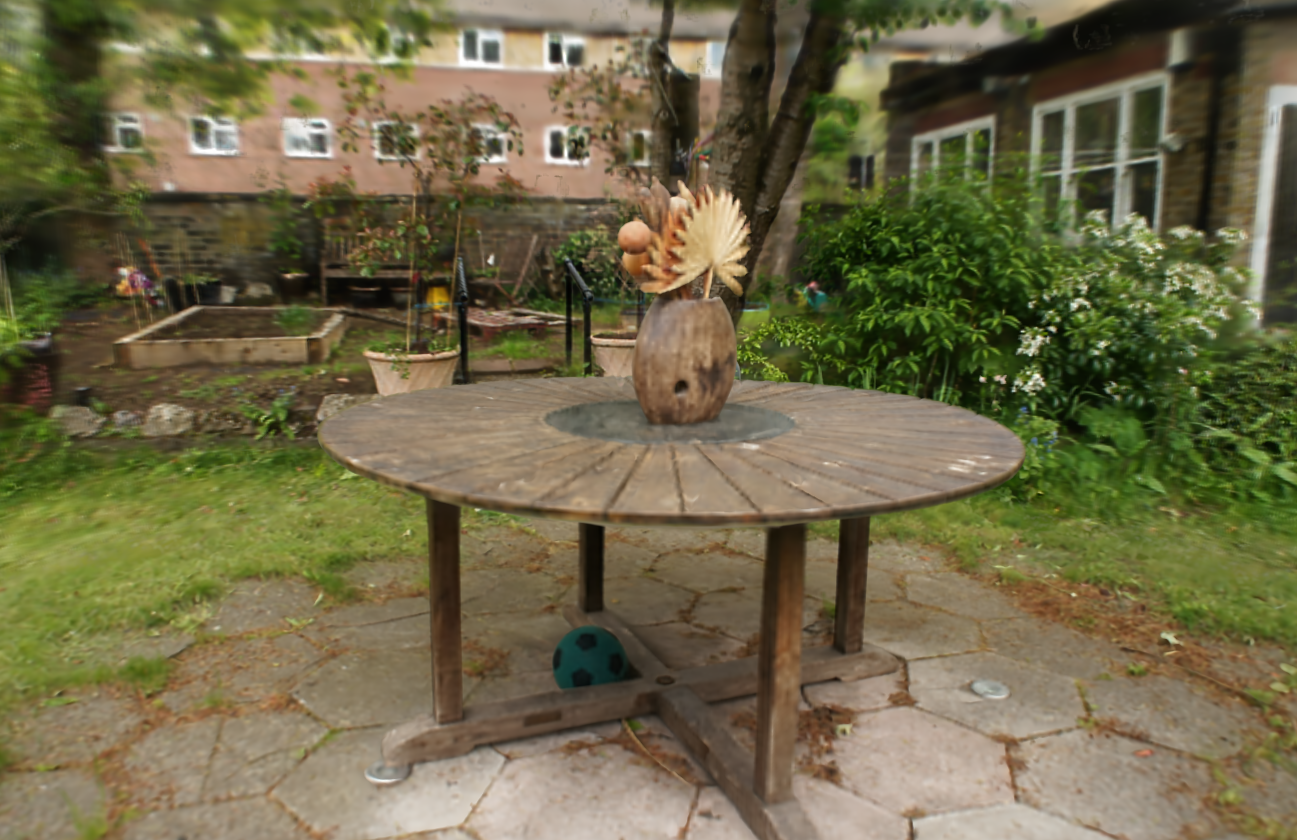}{0.6}{1.95}{0.55cm}{0.55cm}{1cm}{\mytmplen}{3}{red}\\
		\includegraphics[width=\mytmplen]{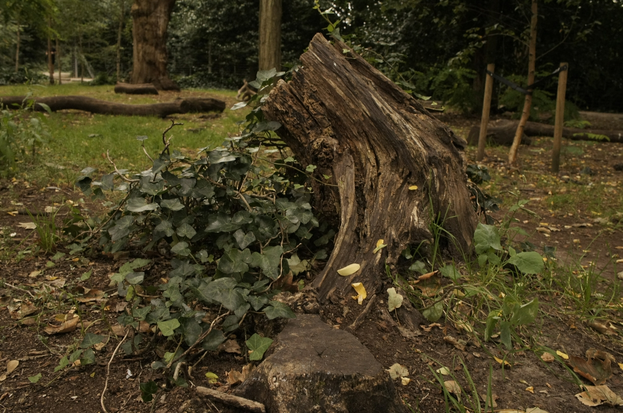} &
		\includegraphics[width=\mytmplen]{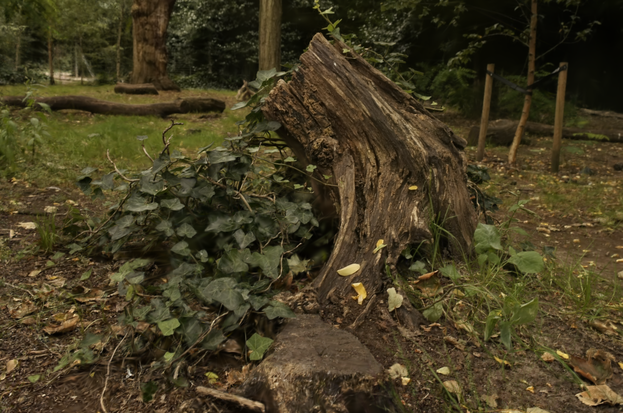} &
		\includegraphics[width=\mytmplen]{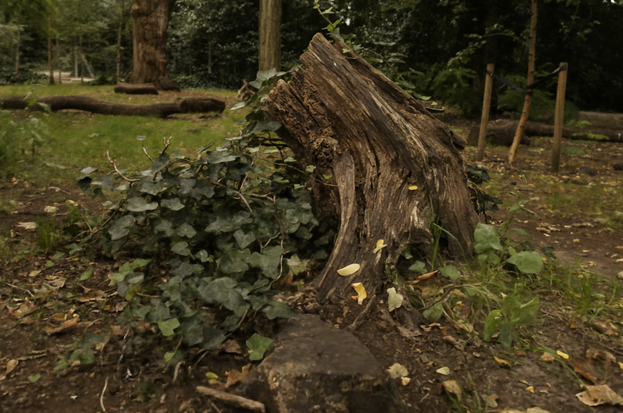} &
		\includegraphics[width=\mytmplen]{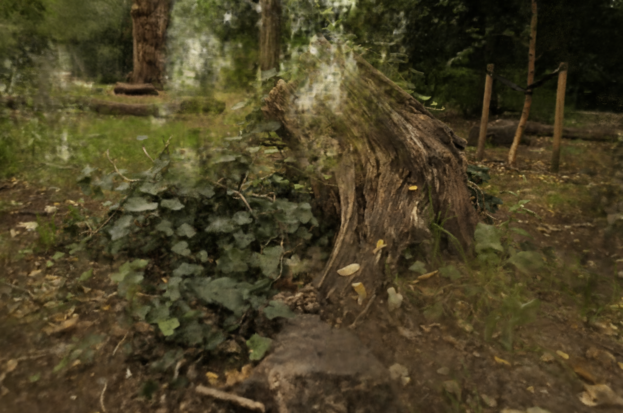} &
		\includegraphics[width=\mytmplen]{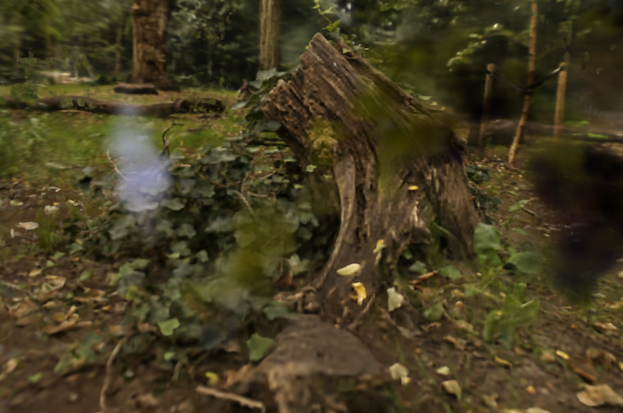} 
		\\
		\includegraphics[width=\mytmplen]{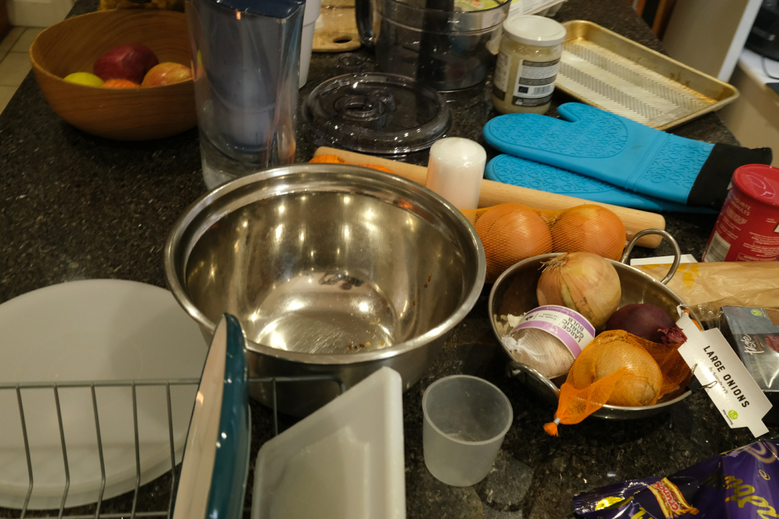} &
		\includegraphics[width=\mytmplen]{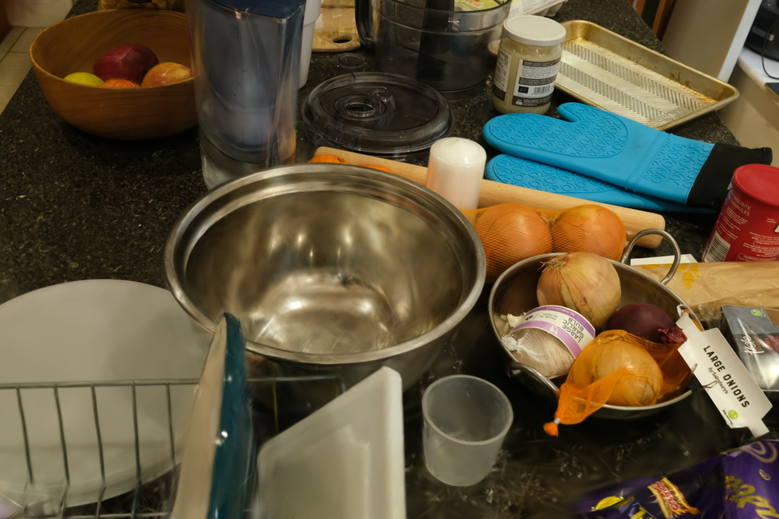} &
 		\begin{tikzpicture}
			\node[anchor=south west,inner sep=0] (image) at (0,0) {\includegraphics[width=\mytmplen]{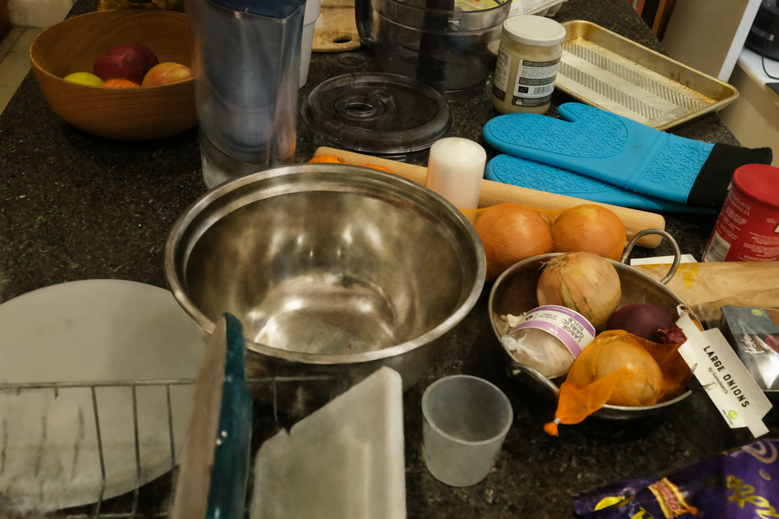}};
			\begin{scope}[x={(image.south east)},y={(image.north west)}]
				\draw[red,ultra thick,->] (0.07,0.43) -- (0.07,0.23);
			\end{scope}
		\end{tikzpicture} &
		\includegraphics[width=\mytmplen]{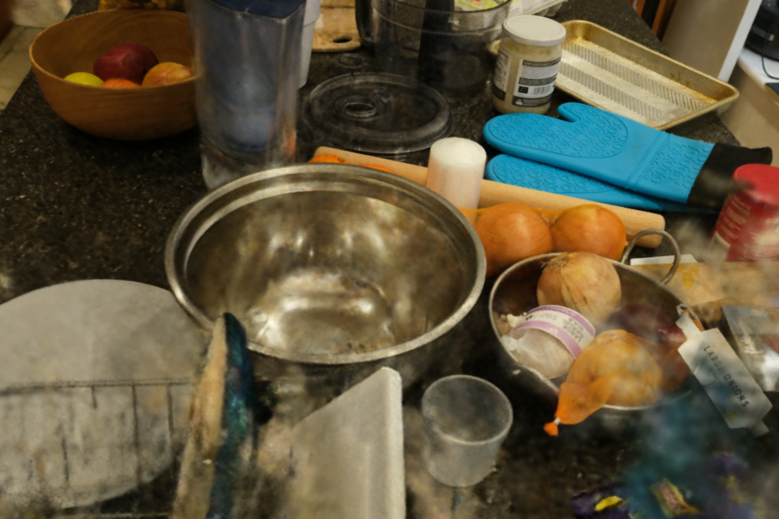} &
		\includegraphics[width=\mytmplen]{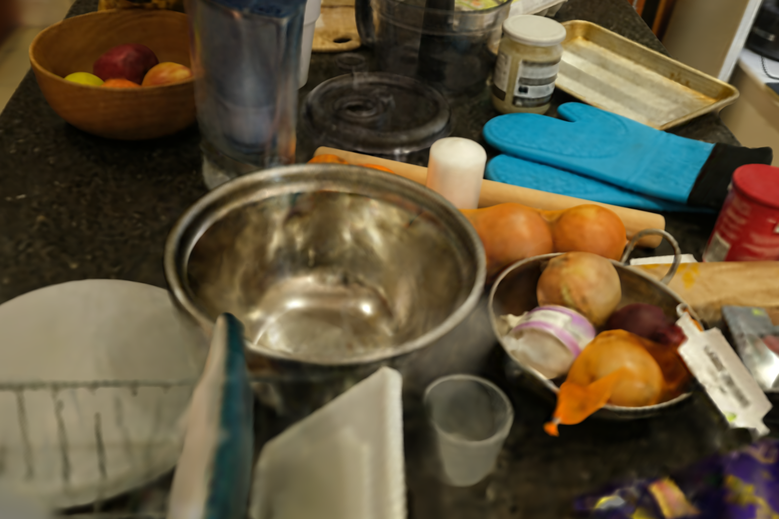} 
		\\
		\includegraphics[width=\mytmplen]{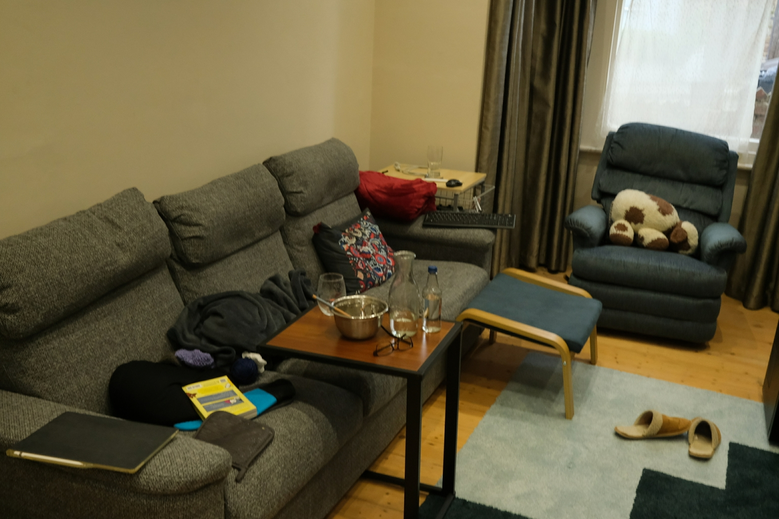} &
		\includegraphics[width=\mytmplen]{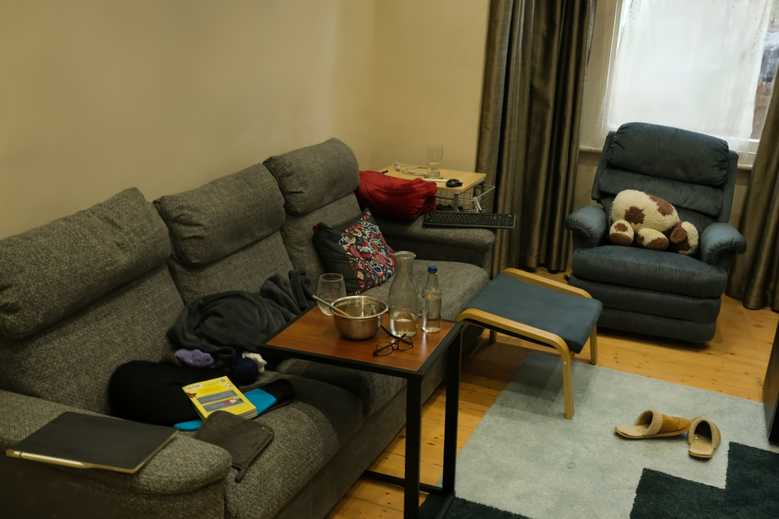} &
		\includegraphics[width=\mytmplen]{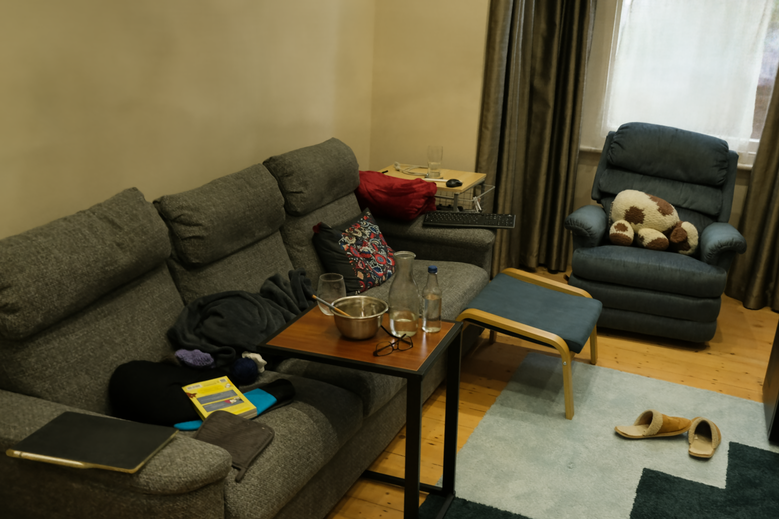} &
		\includegraphics[width=\mytmplen]{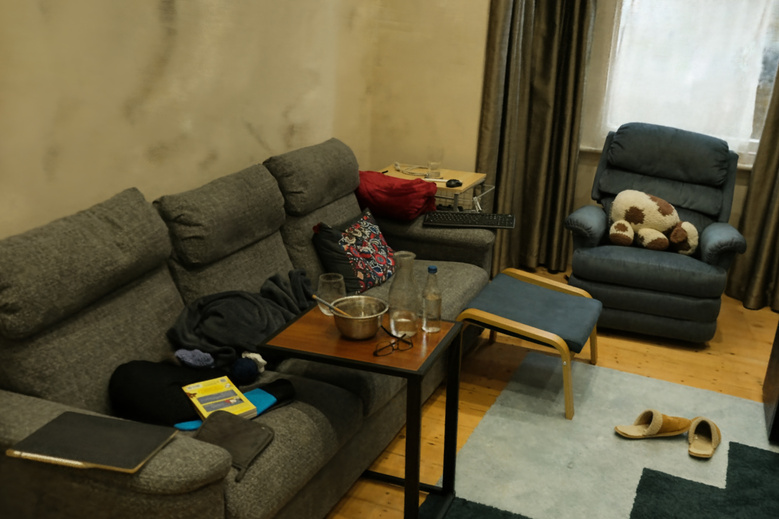} &
		\includegraphics[width=\mytmplen]{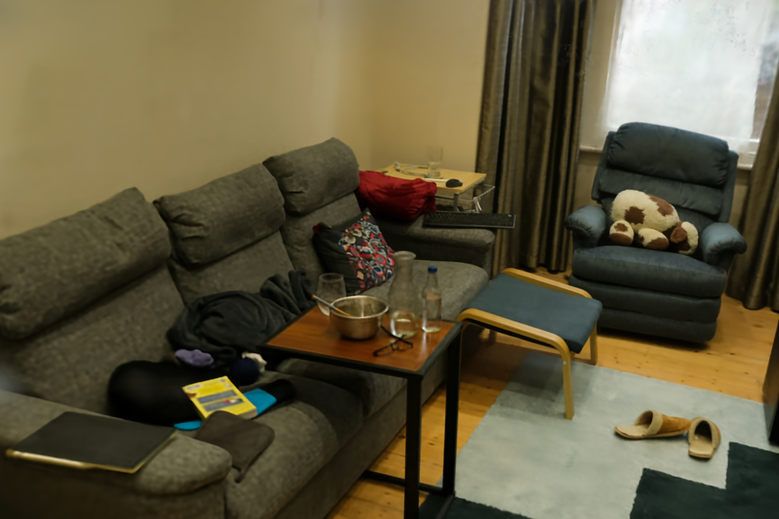} \\
		\zoomin{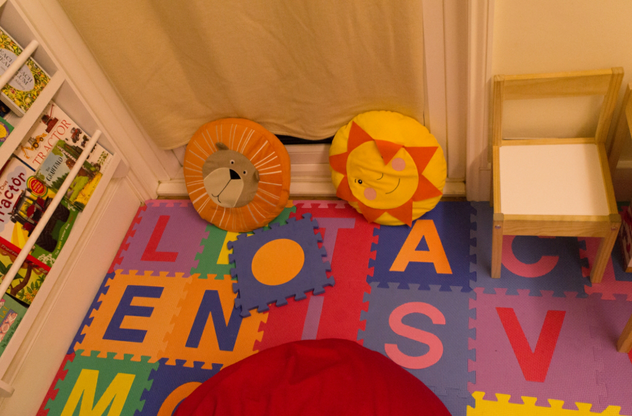}{1.2}{1.4}{0.55cm}{0.55cm}{1cm}{\mytmplen}{3}{green}&
		\zoomin{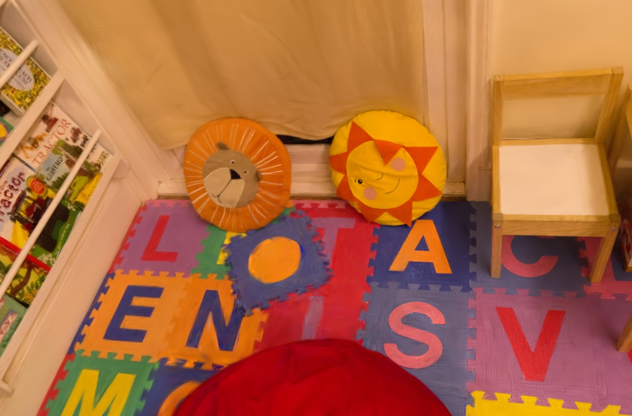}{1.2}{1.4}{0.55cm}{0.55cm}{1cm}{\mytmplen}{3}{green}&
		\zoomin{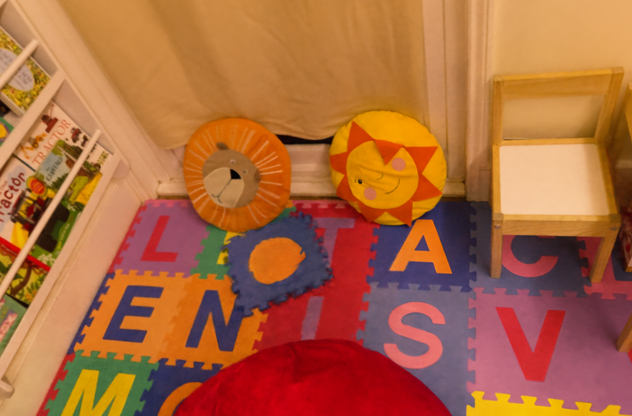}{1.2}{1.4}{0.55cm}{0.55cm}{1cm}{\mytmplen}{3}{green}&
		\zoomin{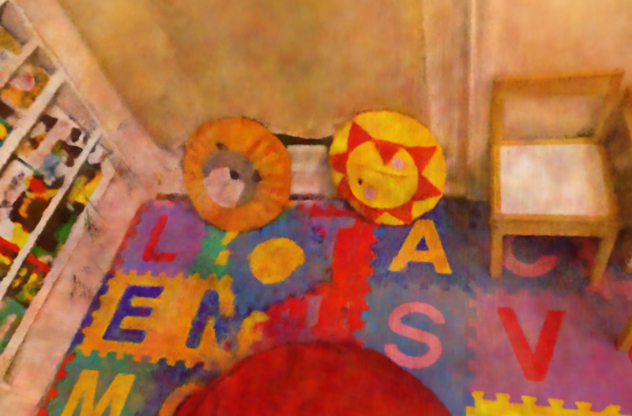}{1.2}{1.4}{0.55cm}{0.55cm}{1cm}{\mytmplen}{3}{green}& 
		\zoomin{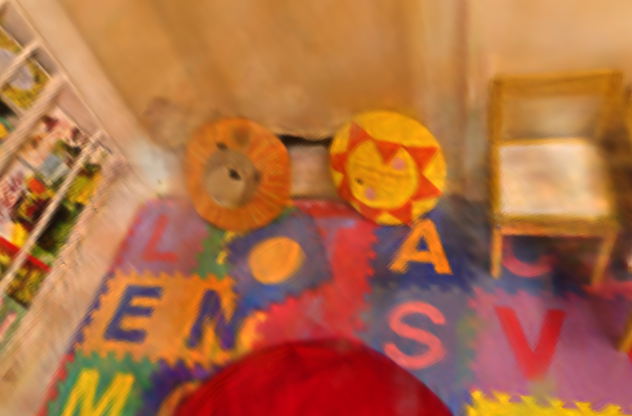}{1.2}{1.4}{0.55cm}{0.55cm}{1cm}{\mytmplen}{3}{green}
\\
		\includegraphics[width=\mytmplen]{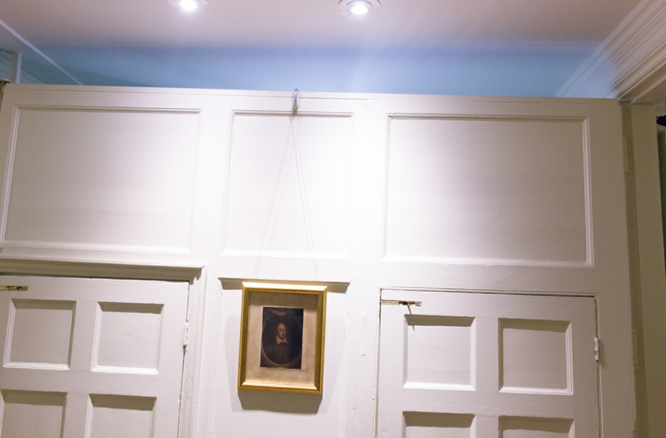} &
		\includegraphics[width=\mytmplen]{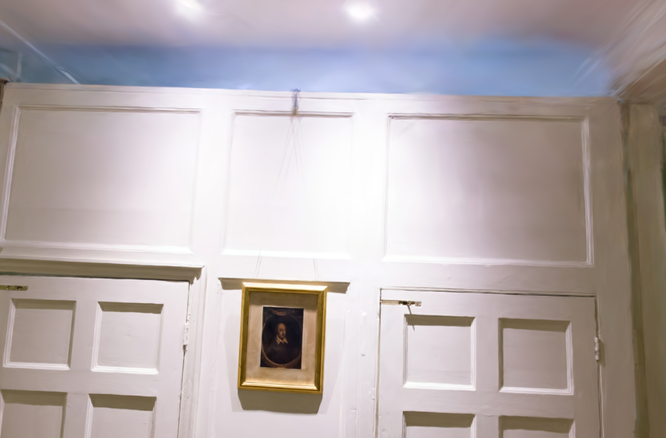} &
		\includegraphics[width=\mytmplen]{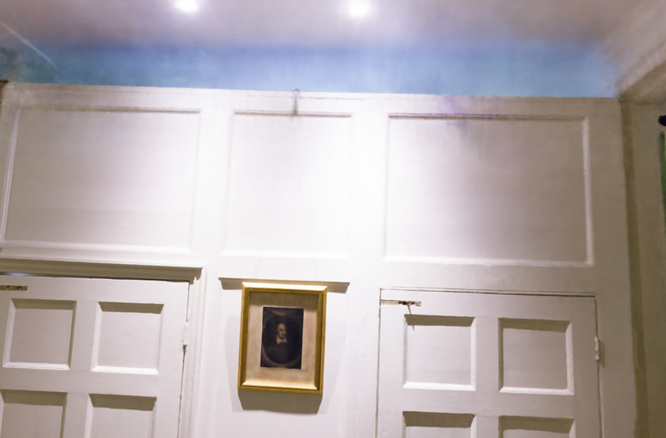} &
		\includegraphics[width=\mytmplen]{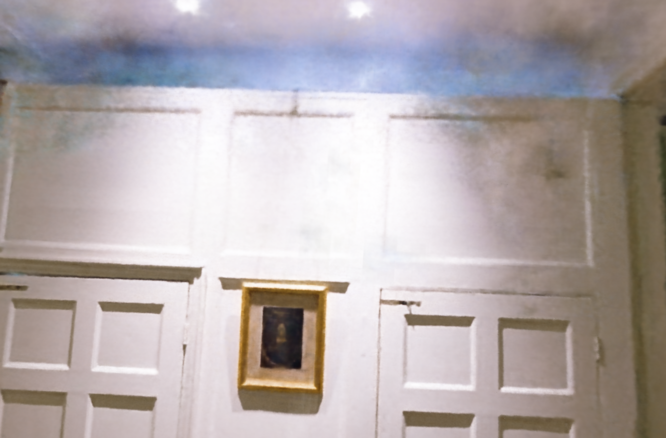} & \includegraphics[width=\mytmplen]{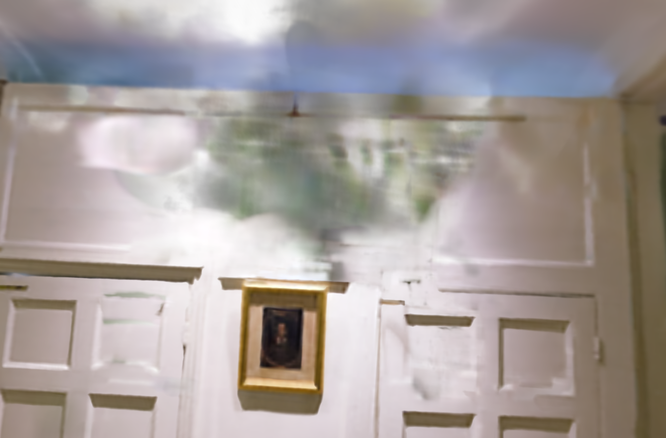} 
\\
		\includegraphics[width=\mytmplen]{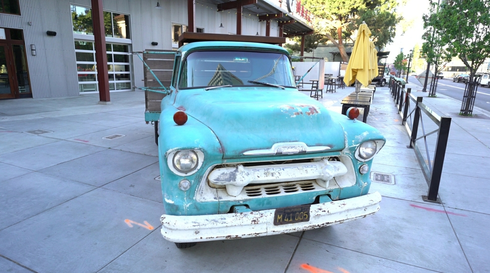} &
		\includegraphics[width=\mytmplen]{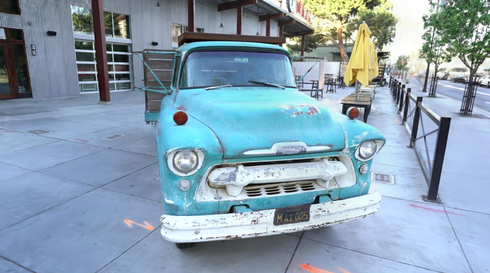} &
		 		\begin{tikzpicture}
			\node[anchor=south west,inner sep=0] (image) at (0,0) {\includegraphics[width=\mytmplen]{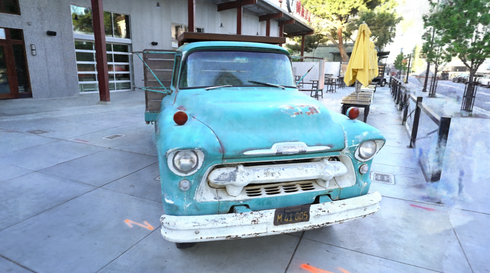}};
			\begin{scope}[x={(image.south east)},y={(image.north west)}]
				\draw[red,ultra thick,->] (0.87,0.11) -- (0.87,0.31);
			\end{scope}
		\end{tikzpicture} &
		\includegraphics[width=\mytmplen]{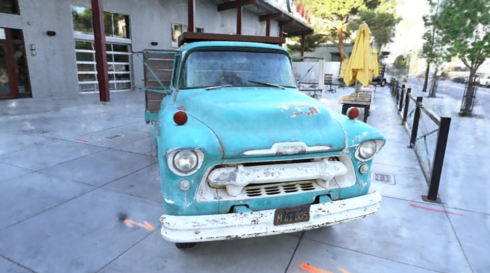} &
		\includegraphics[width=\mytmplen]{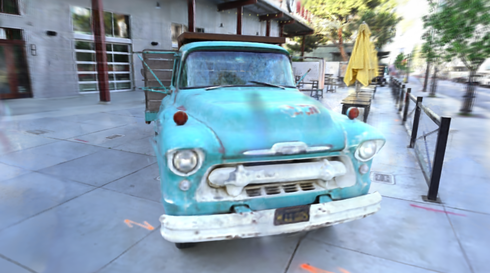} 
\\
		\includegraphics[width=\mytmplen]{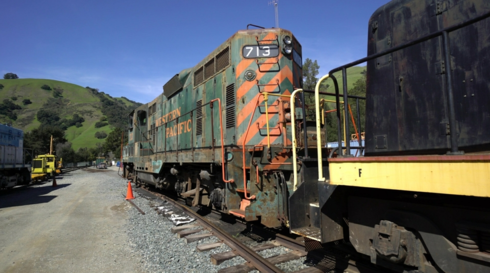} &
		\includegraphics[width=\mytmplen]{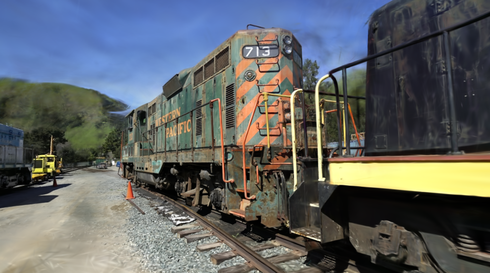} &
		\includegraphics[width=\mytmplen]{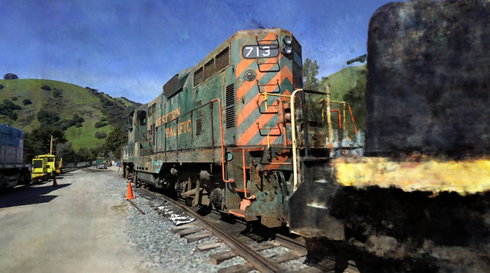} &
		\includegraphics[width=\mytmplen]{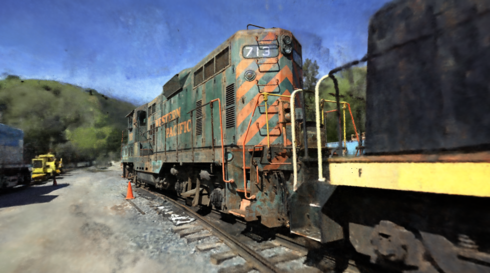} &
		\includegraphics[width=\mytmplen]{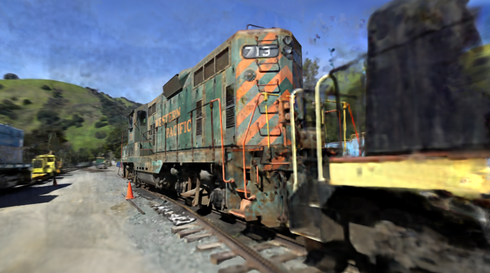} \\
	\end{tabular}
    \vspace{-.3cm} %
	\caption{
		\label{fig:comparisons}
		We show comparisons of ours to previous methods and the corresponding ground truth images from \CORRECTION{paths not in the input views}{held-out test views}. The scenes are, from the top down:
		\textsc{Bicycle}, \textsc{Garden}, \textsc{Stump}, \textsc{Counter} and \textsc{Room} from the Mip-NeRF360 dataset; \textsc{Playroom}, \textsc{DrJohnson} from the Deep Blending dataset~\cite{hedman2018deep} and \textsc{Truck} and \textsc{Train} from Tanks\&Temples. \ADDITION{Non-obvious differences in quality highlighted by arrows/insets.}
	}
\end{figure*}

\begin{table*}[!h]
	\caption{
		\label{tab:comparisons} {\CORRECTION{Quantitative evaluation of our method compared to previous work, computed over our five scenes, on separate paths captured specifically for evaluation, and separate from the input views.}{Quantitative evaluation of our method compared to previous work, computed over three datasets. Results marked with dagger $\dagger$ have been directly adopted from the original paper, all others were obtained in our own experiments.}}
	}
	\small
	\scalebox{0.78}{
		\begin{tabular}{l|cccccc|cccccc|cccccc}
			
			Dataset & \multicolumn{6}{c|}{Mip-NeRF360}  & \multicolumn{6}{c|}{Tanks\&Temples} & \multicolumn{6}{c}{Deep Blending}\\
			Method|Metric
			& $SSIM^\uparrow$   & $PSNR^\uparrow$    & $LPIPS^\downarrow$  & Train & FPS & Mem 
			& $SSIM^\uparrow$   & $PSNR^\uparrow$    & $LPIPS^\downarrow$  & Train & FPS & Mem 
			& $SSIM^\uparrow$   & $PSNR^\uparrow$    & $LPIPS^\downarrow$  & Train & FPS & Mem \\
			\hline 
			Plenoxels& 0.626 & 23.08 & 0.463 & 25m49s & 6.79 & 2.1GB & 0.719 & 21.08 & 0.379 & 25m5s & 13.0 & 2.3GB & 0.795 & 23.06 & 0.510 & 27m49s & 11.2 & 2.7GB \\
			INGP-Base& 0.671 & 25.30 & 0.371 & 5m37s & 11.7 & 13MB & 0.723 & 21.72 & 0.330 & 5m26s & 17.1 & 13MB & 0.797 & 23.62 & 0.423 & 6m31s & 3.26 & 13MB \\ 
			INGP-Big& 0.699 & 25.59 & 0.331 & 7m30s & 9.43 & 48MB & 0.745 & \cellcolor{yellow!40}21.92 & 0.305 & 6m59s & 14.4 & 48MB & 0.817 & 24.96 & 0.390 & 8m & 2.79 & 48MB \\ 
			M-NeRF360&\cellcolor{orange!40}0.792$^\dagger$ & \cellcolor{red!40} 27.69$^\dagger$ & \cellcolor{orange!40}0.237$^\dagger$ & 48h & 0.06 & 8.6MB  & \cellcolor{yellow!40}0.759 & \cellcolor{orange!40}22.22 & \cellcolor{orange!40}0.257 & 48h & 0.14 & 8.6MB & \cellcolor{orange!40}0.901 & \cellcolor{orange!40}29.40 &  \cellcolor{orange!40}0.245 & 48h & 0.09 & 8.6MB\\
			Ours-7K& \cellcolor{yellow!40}0.770 & \cellcolor{yellow!40}25.60 & \cellcolor{yellow!40}0.279 & 6m25s & 160 & 523MB & \cellcolor{orange!40}0.767 & 21.20 & \cellcolor{yellow!40}0.280 & 6m55s & 197 & 270MB & \cellcolor{yellow!40}0.875 &\cellcolor{yellow!40}27.78 & \cellcolor{yellow!40}0.317 & 4m35s & 172 & 386MB \\
			Ours-30K& \cellcolor{red!40}0.815 &  \cellcolor{orange!40}27.21 & \cellcolor{red!40}0.214 & 41m33s & 134 & 734MB & \cellcolor{red!40}0.841 & \cellcolor{red!40}23.14 & \cellcolor{red!40}0.183 & 26m54s & 154 & 411MB & \cellcolor{red!40}0.903 & \cellcolor{red!40}29.41 & \cellcolor{red!40}0.243 & 36m2s & 137 & 676MB\\
			
		\end{tabular}
	}
\end{table*}

\begin{figure}[h]
	\includegraphics[width=\linewidth]{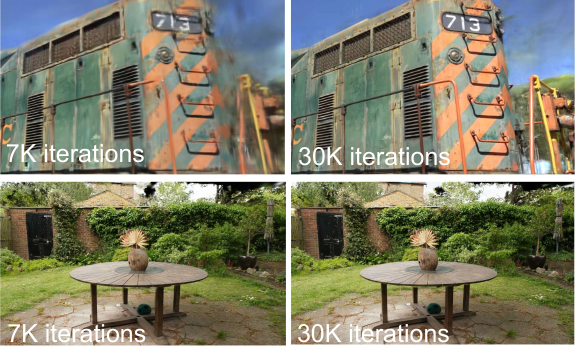} 
	\caption{
		\label{fig:5vs40min}
		For some scenes (above) we can see that even at 7K iterations ($\sim$5min for this scene), our method has captured the train quite well. At 30K iterations ($\sim$35min) the background artifacts have been reduced significantly. For other scenes (below), the difference is barely visible; 7K iterations ($\sim$8min) is already very high quality.
	}
\end{figure}

\section{Implementation, results and evaluation}

We next discuss some details of implementation, present results and the evaluation of our algorithm compared to previous work and ablation studies.

\begin{table}[!h]
	\caption{PSNR scores for Synthetic NeRF, we start with 100K randomly initialized points. \ADDITION{Competing} metrics extracted from respective papers. }
	\scalebox{0.7}{
		\centering
		\begin{tabular}{l|ccccccccc}
			~ & Mic & Chair & Ship & Materials & Lego & Drums & Ficus & Hotdog & Avg.  \\ \hline
			Plenoxels & 33.26 & 33.98 & 29.62 & 29.14 & 34.10 & 25.35 & 31.83 & 36.81 & 31.76  \\
			INGP-Base & \cellcolor{orange!40}36.22 & 35.00 & \cellcolor{red!40}31.10 & \cellcolor{yellow!40}29.78 & \cellcolor{red!40}36.39 & \cellcolor{yellow!40}26.02 & \cellcolor{yellow!40}33.51 & \cellcolor{yellow!40}37.40 & \cellcolor{yellow!40}33.18  \\ 
			Mip-NeRF & \cellcolor{red!40}36.51 & \cellcolor{yellow!40}35.14 & 30.41 & \cellcolor{red!40}30.71 & \cellcolor{yellow!40}35.70 & 25.48 & 33.29 & \cellcolor{orange!40}37.48 & 33.09  \\ 
			Point-NeRF & \cellcolor{yellow!40}35.95 & \cellcolor{orange!40}35.40 & \cellcolor{orange!40}30.97 & 29.61 & 35.04 & \cellcolor{orange!40}26.06 & \cellcolor{red!40}36.13 & 37.30 & \cellcolor{orange!40}33.30  \\  
			Ours-30K & 35.36 & \cellcolor{red!40}35.83 & \cellcolor{yellow!40}30.80 & \cellcolor{orange!40}30.00 & \cellcolor{orange!40}35.78 & \cellcolor{red!40}26.15 & \cellcolor{orange!40}34.87 & \cellcolor{red!40}37.72 & \cellcolor{red!40}33.32  \\
		\end{tabular}
		\label{tab:results_synthetic}
	}
\end{table}
\subsection{Implementation}
\label{sec:impl}

We implemented our method in Python using the PyTorch framework and wrote custom CUDA kernels for rasterization that are extended versions of previous methods~\cite{kopanas21}, and use the NVIDIA CUB sorting routines for the fast Radix sort~\cite{merrill2010revisiting}.
We also built an interactive viewer using the open-source SIBR~\cite{sibr2020}, used for interactive viewing. \ADDITION{We used this implementation to measure our achieved frame rates}.
The source code and all our data are available at: \textcolor{blue}{\url{https://repo-sam.inria.fr/fungraph/3d-gaussian-splatting/}}

\paragraph{Optimization Details}
For stability, we ``warm-up'' the computation in lower resolution. 
Specifically, we start the optimization using 4 times smaller image resolution and we upsample twice after 250 and 500 iterations. 

SH coefficient optimization is sensitive to the lack of angular information. For typical ``NeRF-like''
captures where a central object is observed by photos taken in the entire hemisphere around it, the optimization works well. However, if the capture has angular regions missing (e.g., when capturing the corner of a scene, or performing an ``inside-out''~\cite{HRDB16} capture) completely incorrect values for the zero-order component of the SH (i.e., the base or diffuse color) can be produced by the optimization. 
To overcome this problem we start by optimizing only the zero-order component, and then introduce one band of the SH after every 1000 iterations until all 4 bands of SH are represented. %

\begin{table*}[!h]
	\caption{PSNR Score for ablation runs. \ADDITION{For this experiment, we manually downsampled high-resolution versions of each scene's input images to the established rendering resolution of our other experiments. Doing so reduces random artifacts (e.g., due to JPEG compression in the pre-downscaled Mip-NeRF360 inputs).}}
	\centering
	\begin{tabular}{l|ccc|ccc|cc}
		~ & Truck-5K & Garden-5K & Bicycle-5K & Truck-30K & Garden-30K & Bicycle-30K & Average-5K & Average-30K \\ \hline
		Limited-BW  & 14.66 & 22.07 & 20.77 & 13.84 & 22.88 & 20.87 & 19.16 & 19.19 \\
		Random Init & 16.75 & 20.90 & 19.86 & 18.02 & 22.19 & 21.05 & 19.17 & 20.42 \\ 
		No-Split    & 18.31 & 23.98 & 22.21 & 20.59 & 26.11 & 25.02 & 21.50 & 23.90 \\ 
		No-SH       & \cellcolor{yellow!40}22.36 & 25.22 & \cellcolor{orange!40}22.88 & \cellcolor{yellow!40}24.39 & 26.59 & \cellcolor{yellow!40}25.08 & \cellcolor{yellow!40}23.48 & \cellcolor{yellow!40}25.35 \\ 
		No-Clone    & 22.29 & \cellcolor{orange!40}25.61 & 22.15 & \cellcolor{red!40}24.82 & \cellcolor{orange!40}27.47 & \cellcolor{orange!40}25.46 & 23.35 & \cellcolor{orange!40}25.91 \\  
		Isotropic   & \cellcolor{orange!40}22.40 & \cellcolor{yellow!40}25.49 & \cellcolor{yellow!40}22.81 & 23.89 & \cellcolor{yellow!40}27.00 & 24.81 & \cellcolor{orange!40}23.56 & 25.23 \\  
		Full        & \cellcolor{red!40}22.71 & \cellcolor{red!40}25.82 & \cellcolor{red!40}23.18 & \cellcolor{orange!40}24.81 &\cellcolor{red!40} 27.70 & \cellcolor{red!40}25.65 & \cellcolor{red!40}23.90 & \cellcolor{red!40}26.05 \\  
		
	\end{tabular}
	\label{tab:ablation_table}
\end{table*}

\subsection{Results and Evaluation}

\paragraph{Results.}
We tested our algorithm on a total of \CORRECTION{11}{13} real scenes taken from previously published datasets and the synthetic Blender dataset ~\cite{mildenhall2020nerf}. 
In particular, we tested our approach on the full set of scenes presented in Mip-Nerf360~\cite{barron2022mipnerf360}, which is the current state of the art in NeRF rendering quality, two scenes from the Tanks\&Temples dataset \shortcite{knapitsch2017tanks} and two scenes provided by Hedman et al.~\cite{hedman2018deep}. 
The scenes we chose have very different capture styles, and cover both bounded indoor scenes and large unbounded outdoor environments.
\CORRECTION{We use the same parameter values for all our runs, except for the learning rate for covariance: for all scenes, it is 0.001 except \textsc{Train}, \textsc{DrJohnson} and \textsc{Playroom} (see Fig.~\mbox{\ref{fig:comparisons}}) where it is divided by three because the capture style is much less structured.}{We use the same hyperparameter configuration for all experiments in our evaluation.} All results are reported running on an A6000 GPU\ADDITION{, except for the Mip-NeRF360 method (see below)}.

In supplemental, we show a rendered video path for a selection of scenes that contain views far from the input photos.

\paragraph{Real-World Scenes.}
In terms of quality, the current state-of-the-art is Mip-Nerf360~\cite{barron2021mipnerf}. We compare against this method as a quality benchmark. We also compare against two of the most recent fast NeRF methods: InstantNGP~\cite{mueller2022instant} and Plenoxels~\cite{plenoxels}. 

We use a train/test split for datasets, using the methodology suggested by Mip-NeRF360, taking every 8th photo for test, for consistent and meaningful comparisons to generate the error metrics, using the standard PSNR, L-PIPS, and SSIM metrics used most frequently in the literature; please see \CORRECTION{Tab.}{Table}~\ref{tab:comparisons}. \CORRECTION{Note that we ran the improved version of Mip-NeRF360 code which gives slightly better numbers than in the original publication.}{All numbers in the table are from our own runs of the author's code for all previous methods, except for those of Mip-NeRF360 on their dataset, in which we copied the numbers from the original publication to avoid confusion about the current SOTA. For the images in our
figures, we used our own run of Mip-NeRF360: the numbers for these runs are in Appendix~\ref{sec:appd}.}
We also show the average training time, rendering speed, and memory used to store optimized parameters. We report results for a basic configuration of InstantNGP (Base) that run for 35K iterations as well as a slightly larger network suggested by the authors (Big), and two configurations, 7K and 30K iterations for ours. 
We show the difference in visual quality for our two configurations in Fig.~\ref{fig:5vs40min}. In many cases, quality at 7K iterations is already quite good.

The training times vary over datasets and we report them separately. Note that image resolutions also vary over datasets.
In \CORRECTION{supplemental}{the project website}, we provide all the renders of test views we used to compute the statistics for all the methods (ours and previous work) on \CORRECTION{\textsc{bicycle}, \textsc{Truck} and \textsc{Playroom}}{all scenes.} Note that we kept the native input resolution for all renders.

The table shows that our fully converged model achieves quality that is on par and sometimes slightly better than the SOTA Mip-NeRF360 method; note that on the same hardware, their average training time was 48 hours\footnote{We trained Mip-NeRF360 on a 4-GPU A100 node for 12 hours, equivalent to 48 hours on a single GPU. Note that A100's are faster than A6000 GPUs.}, compared to our 35-45min, and their rendering time is 10s/frame. We achieve comparable quality to InstantNGP and Plenoxels after 5-10m of training, but additional training time allows us to achieve SOTA quality which is not the case for the other fast methods. For Tanks \& Temples, we achieve similar quality as the basic InstantNGP at a similar training time ($\sim$\CORRECTION{8}{7}min in our case).

We also show visual results of this comparison for a left-out test view for ours and the previous rendering methods selected for comparison in Fig.~\ref{fig:comparisons}; the results of our method are for 30K iterations of training. We see that in some cases even Mip-NeRF360 has remaining artifacts that our method avoids (e.g., blurriness in vegetation -- in \textsc{Bicycle, Stump} -- or on the walls in \textsc{Room}).
In the supplemental video and web page we provide comparisons of paths from a distance. Our method tends to preserve visual detail of well-covered regions even from far away, which is not always the case for previous methods.

\paragraph{Synthetic Bounded Scenes}
In addition to realistic scenes, we also evaluate our approach on the synthetic \emph{Blender} dataset \cite{mildenhall2020nerf}. The scenes in question provide an exhaustive set of views, are limited in size, and provide exact camera parameters. In such scenarios, we can achieve state-of-the-art results even with random initialization: we start training from 100K uniformly random Gaussians inside a volume that encloses the scene bounds. Our approach quickly and automatically prunes them to about 6--10K meaningful Gaussians. The final size of the trained model after 30K iterations reaches about 200--500K Gaussians per scene. We report and compare our achieved PSNR scores with previous methods in \CORRECTION{Tab.}{Table}~\ref{tab:results_synthetic} using a white background for compatibility. Examples can be seen in Fig.~\ref{fig:ablation-aniso} (second image from the left) and in supplemental material. \ADDITION{The trained synthetic scenes rendered at 180--300 FPS.}

\paragraph{Compactness}
In comparison to previous explicit scene representations, the anisotropic Gaussians used in our optimization are capable of modelling complex shapes with a lower number of parameters.
We showcase this by evaluating our approach against the highly compact, point-based models obtained by \cite{zhang2022}.
We start from their initial point cloud which is obtained by space carving with foreground masks %
and optimize until we break even with their reported PSNR scores. This usually happens within 2--4 minutes. We surpass their reported metrics using approximately one-fourth of their point count, resulting in an average model size of 3.8 MB, as opposed to their 9 MB. We note that for this experiment, we only used two degrees of our spherical harmonics, similar to theirs.

\subsection{Ablations}
\label{sec:ablations}

We isolated the different contributions and algorithmic choices we made and constructed a set of experiments to measure their effect. Specifically we test the following aspects of our algorithm: initialization from SfM, our densification strategies, anisotropic covariance, the fact that we allow an unlimited number of splats to have gradients
and use of spherical harmonics.
The quantitative effect of each choice is summarized in \CORRECTION{Tab.}{Table}~\ref{tab:ablation_table}.

\begin{figure}[!h]
	\includegraphics[width=\linewidth]{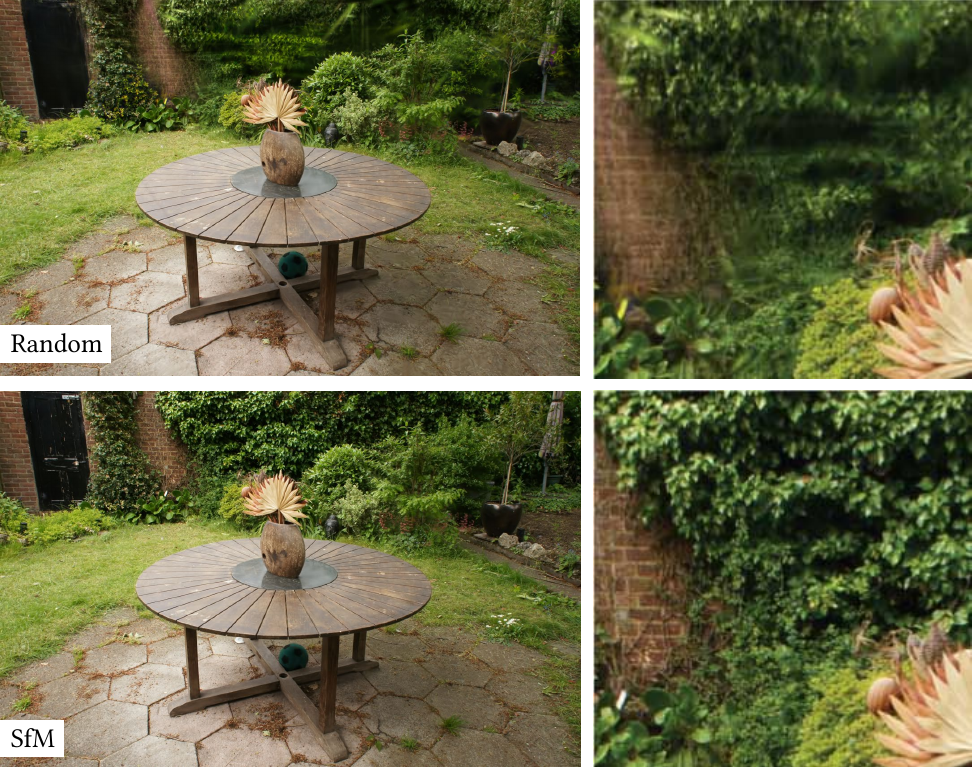}
	\caption{
		\label{fig:random}
		Initialization with SfM points helps. Above: initialization with a random point cloud. Below: initialization using SfM points.
	}
\end{figure}

\paragraph{Initialization from SfM}
We also assess the importance of initializing the 3D Gaussians from the SfM point cloud. 
For this ablation, we uniformly sample a cube with a size equal to three times the extent of the input camera's bounding box. We observe that our method performs relatively well, avoiding complete failure even without the SfM points. Instead, it degrades mainly in the background, see Fig.~\ref{fig:random}. Also in areas not well covered from training views, the random initialization method appears to have more floaters that cannot be removed by optimization. On the other hand, the synthetic NeRF dataset does not have this behavior because it has no background and is well constrained by the input cameras (see discussion above).

\begin{figure}[!h]
	\includegraphics[width=\linewidth]{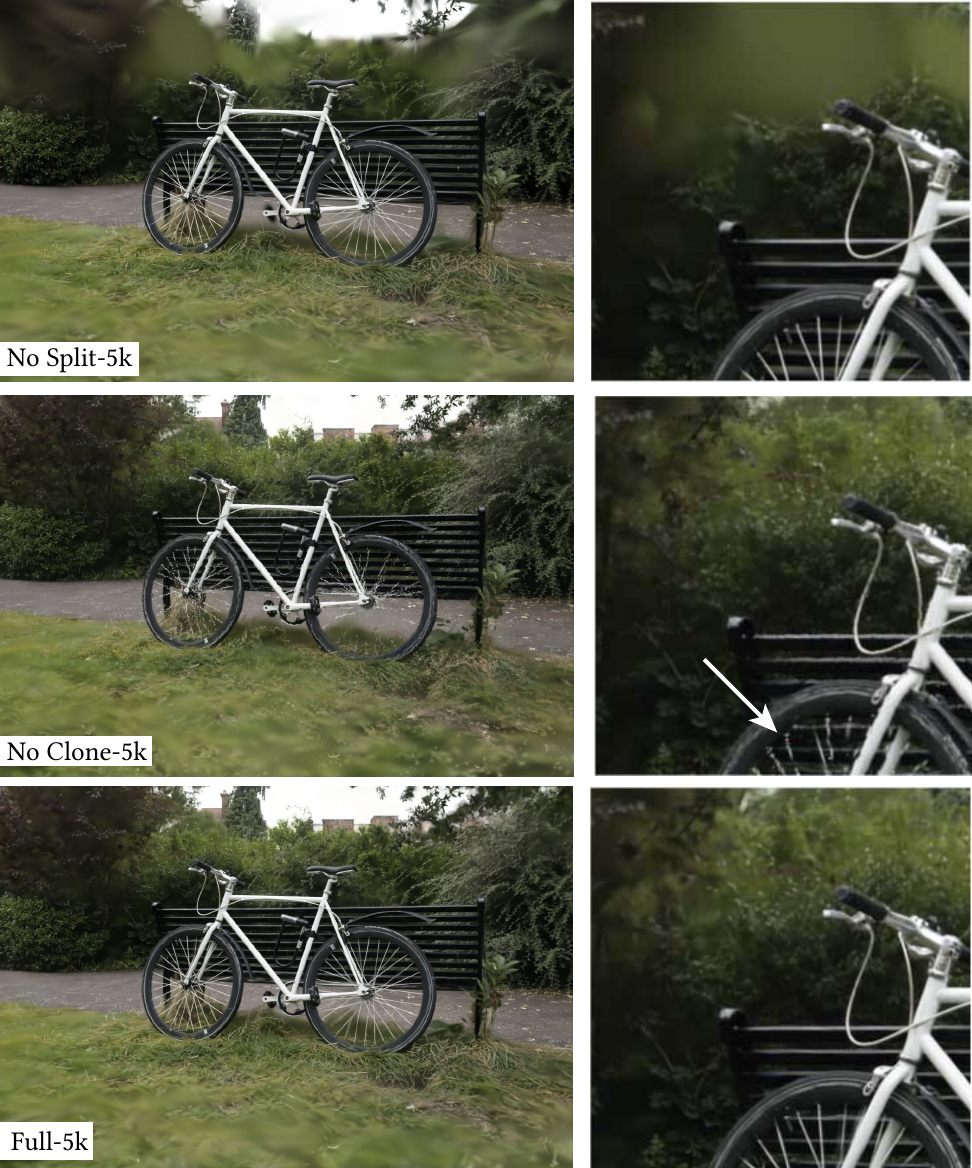}
	\caption{
		\label{fig:densification_ablate}
		Ablation of densification strategy for the two cases "clone" and "split" (Sec.~\ref{sec:opt-dens}).
	}
\end{figure}

\paragraph{Densification.}
We next evaluate our two densification methods, more specifically the clone and split strategy described in Sec.~\ref{sec:opt-dens}. We disable each method separately and optimize using the rest of the method unchanged. Results show that splitting big Gaussians is important to allow good reconstruction of the background as seen in Fig.~\ref{fig:densification_ablate}, while cloning the small Gaussians instead of splitting them allows for a better and faster convergence especially when thin structures appear in the scene. 

\begin{figure}[!h]
	\includegraphics[width=.49\linewidth]{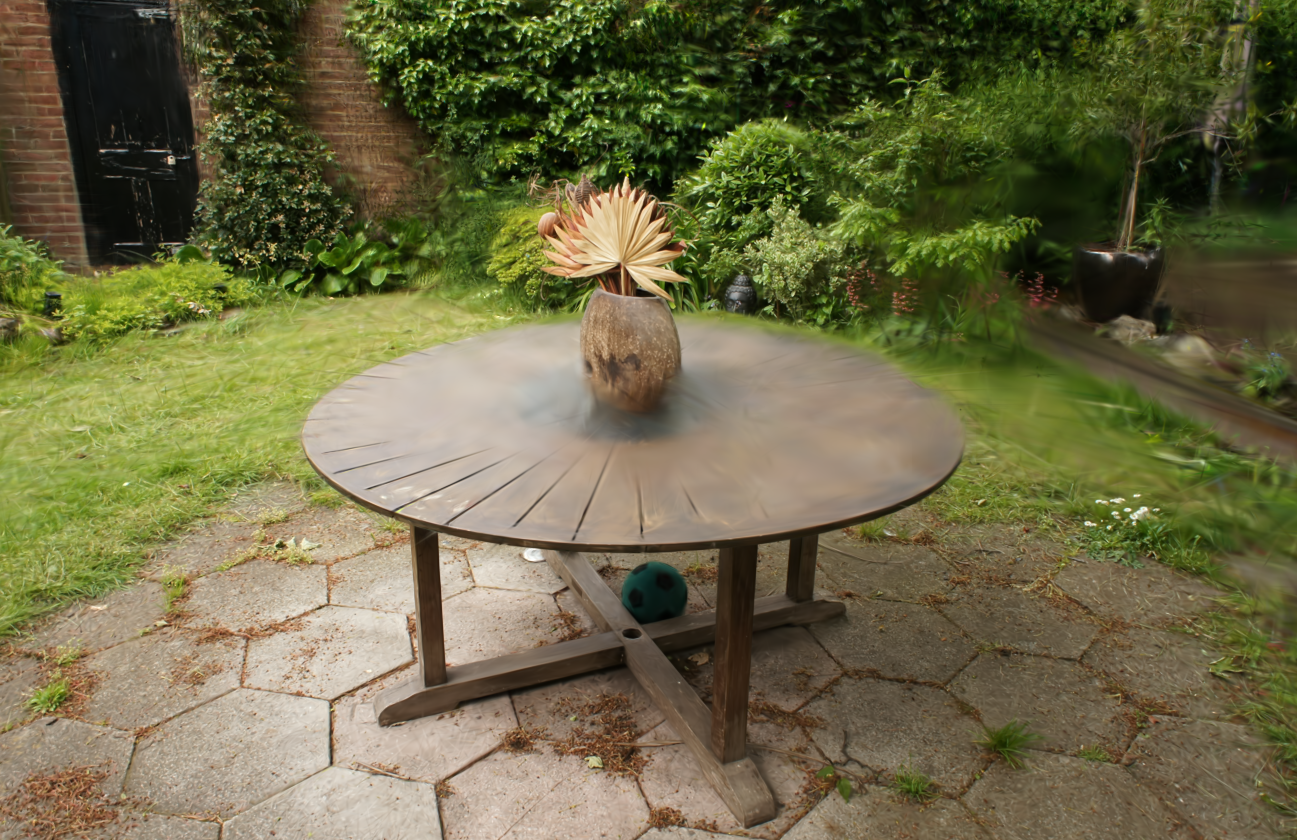}
	\includegraphics[width=.49\linewidth]{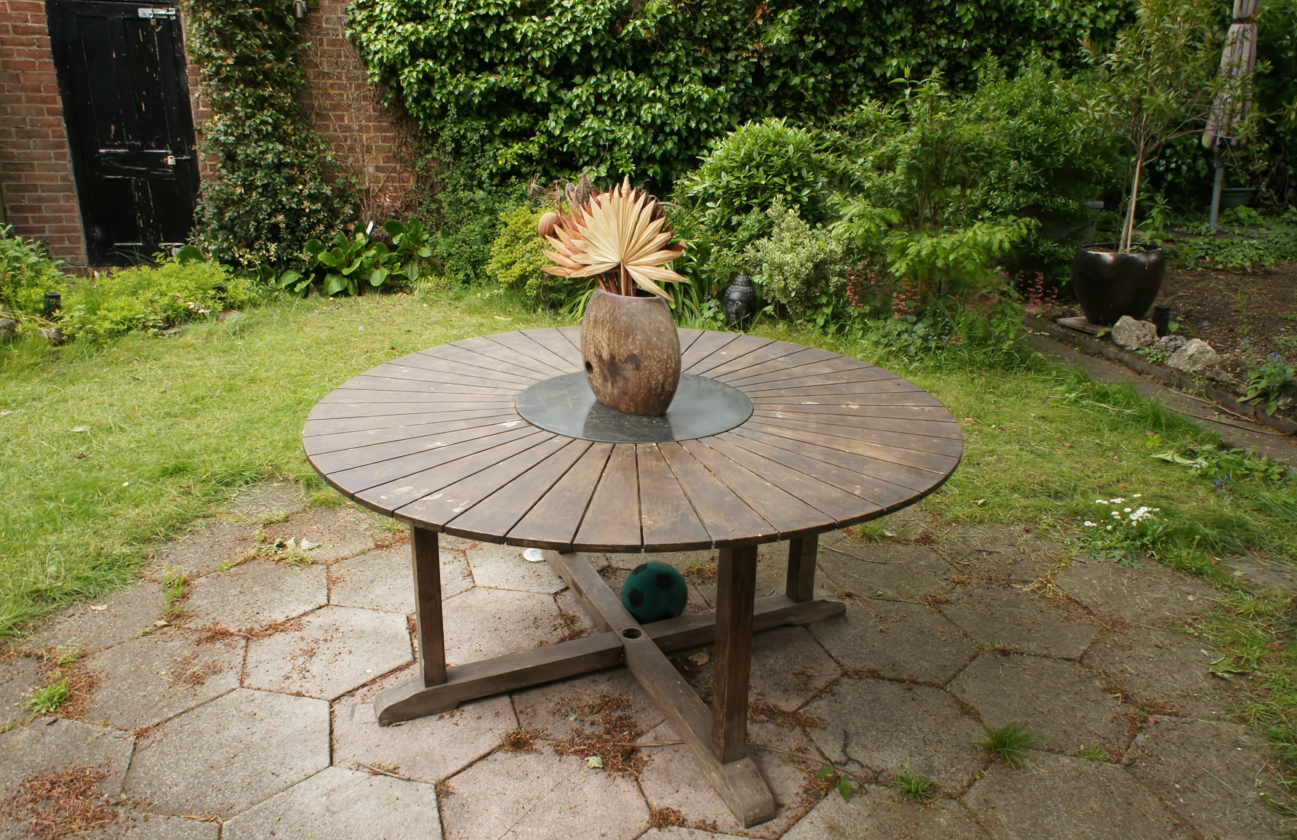}
	\caption{
		If we limit the number of points that receive gradients, the effect on visual quality is significant.
		Left: limit of 10 Gaussians that receive gradients. Right: our full method.
		\label{fig:gradients}
	}
\end{figure}

\paragraph{Unlimited depth complexity of splats with gradients.}
We evaluate if skipping the gradient computation after the $N$ front-most points will give us speed without sacrificing quality, as suggested in ~Pulsar~\cite{Lassner_2021_CVPR}. In this test, we choose N=10, which is two times higher than the default value in ~Pulsar, but it led to unstable optimization because of the severe approximation in the gradient computation. For the \textsc{Truck} scene, quality degraded by 11dB in PSNR (see \CORRECTION{Tab.}{Table}~\ref{tab:ablation_table}, Limited-BW), and the visual outcome is shown in Fig.~\ref{fig:gradients} for \textsc{Garden}.

\begin{figure*}[!h]
	\includegraphics{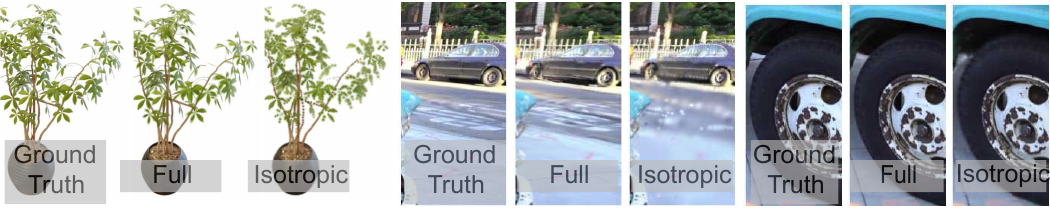}
	\caption{
		\label{fig:ablation-aniso}
		We train scenes with Gaussian anisotropy disabled and enabled. The use of anisotropic volumetric splats enables modelling of fine structures and has a significant impact on visual quality. Note that for illustrative purposes, we restricted \textsc{Ficus} to use no more than 5k Gaussians in both configurations.
	}
\end{figure*}

\paragraph{Anisotropic Covariance.}
An important algorithmic choice in our method is the optimization of the full covariance matrix for the 3D Gaussians. To demonstrate the effect of this choice, we perform an ablation where we remove anisotropy by optimizing a single scalar value that controls the radius of the 3D Gaussian on all three axes. The results of this optimization are presented visually in Fig.~\ref{fig:ablation-aniso}. We observe that the anisotropy significantly improves the quality of the 3D Gaussian's ability to align with surfaces, which in turn allows for much higher rendering quality while maintaining the same number of points.

\paragraph{Spherical Harmonics.}
Finally, the use of spherical harmonics improves our overall PSNR scores since they compensate for the view-dependent effects (\CORRECTION{Tab.}{Table}~\ref{tab:ablation_table}).

\subsection{Limitations}

\begin{figure}[!h]
	\includegraphics[width=0.49\columnwidth]{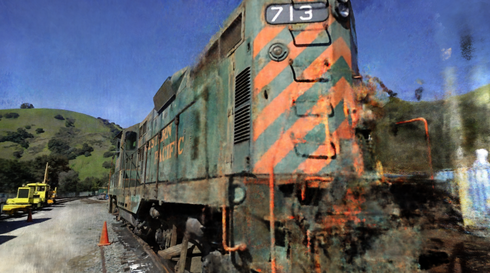}
	\includegraphics[width=0.49\columnwidth]{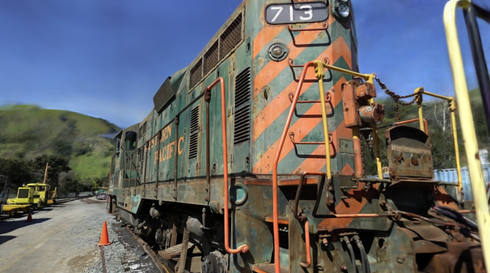}
	\caption{
		\label{fig:limit}
		Comparison of failure artifacts: Mip-NeRF360 has ``floaters'' and grainy appearance (left, foreground), while our method produces coarse, anisoptropic Gaussians resulting in low-detail visuals (right, background). \textsc{Train} scene.
	}
\end{figure}

\begin{figure}[!h]
	\includegraphics[width=0.49\linewidth]{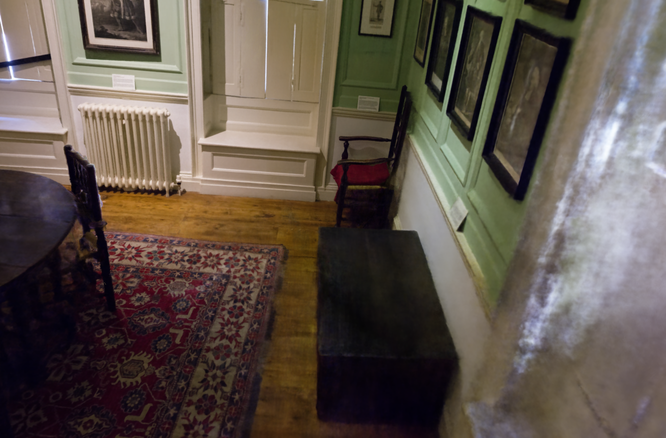}
	\includegraphics[width=0.49\linewidth]{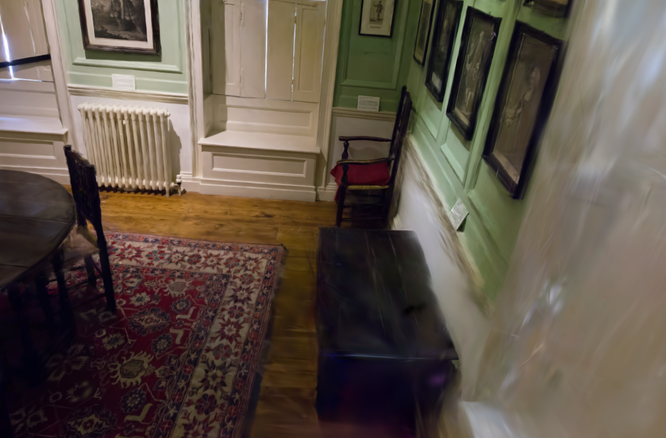}
	\caption{
		In views that have little overlap with those seen during training, our method may produce artifacts (right). Again, Mip-NeRF360 also has artifacts in these cases (left). \textsc{DrJohnson} scene.
		\label{fig:limit-under}
	}
\end{figure}

Our method is not without limitations. 
In regions where the scene is not well observed we have artifacts; in such regions, other methods also struggle (e.g., Mip-NeRF360 in Fig.~\ref{fig:limit}). 
Even though the anisotropic Gaussians have many advantages as described above, 
our method can create elongated artifacts or ``splotchy'' Gaussians (see Fig.~\ref{fig:limit-under}); again previous methods also struggle in these cases.  

We also occasionally have popping artifacts when our optimization creates large Gaussians; this tends to happen in regions with view-dependent appearance. %
One reason for these popping artifacts is the trivial rejection of Gaussians via a guard band in the rasterizer. A more principled culling approach would alleviate these artifacts. \ADDITION{Another factor is our simple visibility algorithm, which can lead to Gaussians suddenly switching depth/blending order. This could be addressed by antialiasing, which we leave as future work.}
Also, we currently do not apply any regularization to our optimization; doing so would help with both the unseen region and popping artifacts. 

\ADDITION{While we used the same hyperparameters for our full evaluation, early experiments show that reducing the position learning rate can be necessary to converge in very large scenes (e.g., urban datasets).}

Even though we are very compact compared to previous point-based approaches, our memory consumption is significantly higher than NeRF-based solutions. \ADDITION{During training of large scenes, peak GPU memory consumption can exceed 20~GB in our unoptimized prototype. However, this figure could be significantly reduced by a careful low-level implementation of the optimization logic (similar to InstantNGP). Rendering the trained scene requires sufficient GPU memory to store the full model (several hundred megabytes for large-scale scenes) and an additional 30--500 MB for the rasterizer, depending on scene size and image resolution. We note that there are many opportunities to further reduce memory consumption of our method.} Compression techniques for point clouds is a well-studied field~\cite{de2016compression}; it would be interesting to see how such approaches could be adapted to our representation.

\section{Discussion and Conclusions}

We have presented the first approach that truly allows real-time, high-quality radiance field rendering, in a wide variety of scenes and capture styles, while requiring training times competitive with the fastest previous methods.

Our choice of a 3D Gaussian primitive preserves properties of volumetric rendering for optimization while directly allowing fast splat-based rasterization. Our work demonstrates that -- contrary to widely accepted opinion -- a continuous representation is \emph{not} strictly necessary to allow fast and high-quality radiance field training.

The majority ($\sim$80\%) of our training time is spent in Python code, since we built our solution in PyTorch to allow our method to be easily used by others. Only the rasterization routine is implemented as optimized CUDA kernels. We expect that porting the remaining optimization entirely to CUDA, as e.g., done in InstantNGP~\cite{mueller2022instant}, could enable significant further speedup for applications where performance is essential.

We also demonstrated the importance of building on real-time rendering principles, exploiting the power of the GPU and speed of software rasterization pipeline architecture. These design choices are the key to performance both for training and real-time rendering, providing a competitive edge in performance over previous volumetric ray-marching.

It would be interesting to see if our Gaussians can be used to perform mesh reconstructions of the captured scene. Aside from practical implications given the widespread use of meshes, this would allow us to better understand where our method stands exactly in the continuum between volumetric and surface representations.

In conclusion, we have presented the first real-time rendering solution for radiance fields, with rendering quality that matches the best expensive previous methods, with training times competitive with the fastest existing solutions.

\begin{acks}
This research was funded by the ERC Advanced grant FUNGRAPH No 788065 \textcolor{blue}{\url{http://fungraph.inria.fr}}. The authors are grateful to Adobe for generous donations, the OPAL infrastructure from Université Côte d’Azur  and for the HPC resources from GENCI–IDRIS (Grant 2022-AD011013409). The authors thank the anonymous reviewers for their valuable feedback, P.\ Hedman and A.\ Tewari for proofreading earlier drafts also T.\ Müller, A.\ Yu and S.\ Fridovich-Keil for helping with the comparisons. 
\end{acks}

\bibliographystyle{ACM-Reference-Format}
\bibliography{points.bib}

\begin{appendix}
	\section{Details of Gradient Computation}
	\label{sec:appa}
	Recall that $\Sigma$/$\Sigma'$ are the world/view space covariance matrices of the Gaussian, $q$ is the rotation, and $s$ the scaling, $W$ is the viewing transformation and $J$ the Jacobian of the affine approximation of the projective transformation.
	We can apply the chain rule to find the derivatives w.r.t.\ scaling and rotation:
	\begin{equation}
		\frac{d\Sigma'}{ds} = \frac{d\Sigma'}{d\Sigma}\frac{d\Sigma}{ds}
	\end{equation}
	and 
	\begin{equation}
		\frac{d\Sigma'}{dq} = \frac{d\Sigma'}{d\Sigma}\frac{d\Sigma}{dq}
	\end{equation}
	Simplifying Eq.~\ref{eq:volume-render} using $U = JW$ and $\Sigma'$ being the (symmetric) upper left $2\times2$ matrix of $U \Sigma U^T$, denoting matrix elements with subscripts, we can find the  partial derivatives
	$
		\frac{\partial \Sigma'}{\partial \Sigma_{ij}} = \left(\begin{smallmatrix}
			U_{1,i}U_{1,j} & U_{1,i} U_{2,j}\\
			U_{1,j} U_{2,i} & U_{2,i}U_{2,j}
		\end{smallmatrix}\right)
	$.
	
	Next, we seek the derivatives $\frac{d\Sigma}{ds}$ and $\frac{d\Sigma}{dq}$. Since $\Sigma = RSS^TR^T$, we can compute $M = RS$ and rewrite $\Sigma = MM^T$. Thus, we can write $\frac{d\Sigma}{ds} = \frac{d\Sigma}{dM} \frac{dM}{ds}$ and $\frac{d\Sigma}{dq} = \frac{d\Sigma}{dM} \frac{dM}{dq}$. Since the covariance matrix $\Sigma$ (and its gradient) is symmetric, the shared first part is compactly found by $\frac{d\Sigma}{dM} = 2M^T$. For scaling, we further have
	$	\frac{\partial M_{i,j}}{\partial s_k} =  \left\{\begin{array}{lr}
		R_{i,k} & \text{if j = k}\\
		0 & \text{otherwise}
	\end{array}\right\}$.
	To derive gradients for rotation, we recall the conversion from a unit quaternion $q$ with real part $q_r$ and imaginary parts $q_i, q_j, q_k$ to a rotation matrix $R$:
	\begin{equation}
		R(q) = 2\begin{pmatrix}
			\frac{1}{2} - (q_j^2 + q_k^2) & (q_i q_j - q_r q_k) & (q_i q_k + q_r q_j)\\
			(q_i q_j + q_r q_k) & \frac{1}{2} - (q_i^2 + q_k^2) & (q_j q_k - q_r q_i)\\
			(q_i q_k - q_r q_j) & (q_j q_k + q_r q_i) & \frac{1}{2} - (q_i^2 + q_j^2)
		\end{pmatrix}
		\label{eq:quat}
	\end{equation}
	As a result, we find the following gradients for the components of $q$:
	\begin{equation}
		\begin{aligned}
			&\frac{\partial M}{\partial q_r} = 2 \left(\begin{smallmatrix}
				0 & -s_y q_k & s_z q_j\\
				s_x q_k & 0 & -s_z q_i\\
				-s_x q_j & s_y q_i & 0
			\end{smallmatrix}\right), 
			&\frac{\partial M}{\partial q_i} = 2\left(\begin{smallmatrix}
				0 & s_y q_j & s_z q_k\\
				s_x q_j & -2 s_y q_i & -s_z q_r\\
				s_x q_k & s_y q_r & -2 s_z q_i
			\end{smallmatrix}\right)
			\\
			&\frac{\partial M}{\partial q_j} = 2\left(\begin{smallmatrix}
				-2 s_x q_j & s_y q_i & s_z q_r\\
				s_x q_i & 0 & s_z q_k\\
				-s_x q_r & s_y q_k & -2s_z q_j
			\end{smallmatrix}\right),
			&\frac{\partial M}{\partial q_k} = 2\left(\begin{smallmatrix}
				-2 s_x q_k & -s_y q_r & s_z q_i\\
				s_x q_r & -2s_y q_k & s_z q_j\\
				s_x q_i & s_y q_j & 0
			\end{smallmatrix}\right)
		\end{aligned}
	\end{equation} 
Deriving gradients for quaternion normalization is straightforward.
	
	\section{Optimization and Densification Algorithm}
	
	Our optimization and densification algorithms are summarized in Algorithm \ref{alg:optimization}.
	\begin{algorithm}[!h]
		\caption{Optimization and Densification\\
		$w$, $h$: width and height of the training images}
		\label{alg:optimization}
		\begin{algorithmic}
			\State $M \gets$ SfM Points	\Comment{Positions}
			\State $S, C, A \gets$ InitAttributes() \Comment{Covariances, Colors, Opacities}
			\State $i \gets 0$	\Comment{Iteration Count}
			
			\While{not converged}
			
			\State $V, \hat{I} \gets$ SampleTrainingView()	\Comment{Camera $V$ and Image}
			\State $I \gets$ Rasterize($M$, $S$, $C$, $A$, $V$)	\Comment{Alg.~\ref{alg:rasterize}}
			
			\State $L \gets Loss(I, \hat{I}) $ \Comment{Loss}
			
			\State $M$, $S$, $C$, $A$ $\gets$ Adam($\nabla L$) \Comment{Backprop \& Step}

			\If{IsRefinementIteration($i$)}
			\ForAll{Gaussians $(\mu, \Sigma, c, \alpha)$ $\textbf{in}$ $(M, S, C, A)$}
			\If{$\alpha < \epsilon$ or IsTooLarge($\mu, \Sigma)$}	\Comment{Pruning}
			\State RemoveGaussian()	
			\EndIf
			\If{$\nabla_p L > \tau_p$} \Comment{Densification}
			\If{$\|S\| > \tau_S$}	\Comment{Over-reconstruction}
			\State SplitGaussian($\mu, \Sigma, c, \alpha$)
			\Else								\Comment{Under-reconstruction}
			\State CloneGaussian($\mu, \Sigma, c, \alpha$)
			\EndIf	
			\EndIf
			\EndFor		
			\EndIf
			\State $i \gets i+1$
			\EndWhile
		\end{algorithmic}
	\end{algorithm}

	\section{Details of the Rasterizer}
	\label{app:raster}

	\paragraph{Sorting.}
	Our design is based on the assumption of a high load of small splats, and we optimize for this by sorting splats once for each frame using radix sort at the beginning.
	We split the screen into 16x16 pixel tiles (or bins).  We create a list of splats per tile by instantiating each splat in each 16$\times$16 tile it overlaps. This results in a moderate increase in Gaussians to process 
	which however is amortized by simpler control flow and high parallelism of optimized GPU Radix sort~\cite{merrill2010revisiting}.
	We assign a key for each splats instance with up to 64 bits where the lower 32 bits encode its projected depth and the higher bits encode the index of the overlapped tile. The exact size of the index depends on how many tiles fit the current resolution. Depth ordering is thus directly resolved for all splats in parallel with a single radix sort. After sorting, we can efficiently produce per-tile lists of Gaussians to process by identifying the start and end of ranges in the sorted array with the same tile ID. This is done in parallel, launching one thread per 64-bit array element to compare its higher 32 bits with its two neighbors. 
	Compared to \cite{Lassner_2021_CVPR}, our rasterization thus completely eliminates sequential primitive processing steps and produces more compact per-tile lists to traverse during the forward pass.
	We show a high-level overview of the rasterization approach in Algorithm~\ref{alg:rasterize}.
	
	\begin{algorithm}
		\caption{GPU software rasterization of 3D Gaussians\\
			$w$, $h$: width and height of the image to rasterize\\
			$M$, $S$: Gaussian means and covariances in world space\\
			$C$, $A$: Gaussian colors and opacities\\
			$V$: view configuration of current camera}
		\label{alg:rasterize}
		\begin{algorithmic}

			\Function{Rasterize}{$w$, $h$, $M$, $S$, $C$, $A$, $V$}
			
			\State CullGaussian($p$, $V$) \Comment{Frustum Culling}
			\State $M', S'$ $\gets$ ScreenspaceGaussians($M$, $S$, $V$) \Comment{Transform}
			\State $T$ $\gets$ CreateTiles($w$, $h$)
			\State $L$, $K$ $\gets$ DuplicateWithKeys($M'$, $T$) \Comment{Indices and Keys}
			\State SortByKeys($K$, $L$)							\Comment{Globally Sort}
			\State $R$ $\gets$ IdentifyTileRanges($T$, $K$)
			\State $I \gets \mathbf{0}$ \Comment{Init Canvas}
			
			\ForAll{Tiles $t$ $\textbf{in}$ $I$}
			\ForAll{Pixels $i$ $\textbf{in}$ $t$}
			
			\State $r \gets$ GetTileRange($R$, $t$)
			
			\State $I[i] \gets$ BlendInOrder($i$, $L$, $r$, $K$, $M'$, $S'$, $C$, $A$)

			\EndFor
			\EndFor

			\Return $I$
			\EndFunction
			
		\end{algorithmic}
	\end{algorithm}

	\ADDITION{\paragraph{Numerical stability.} During the backward pass, we reconstruct the intermediate opacity values needed for gradient computation by repeatedly dividing the accumulated opacity from the forward pass by each Gaussian's $\alpha$. Implemented na\"{i}vely, this process is prone to numerical instabilities (e.g., division by 0). To address this, both in the forward and backward pass, we skip any blending updates with $\alpha < \epsilon$ (we choose $\epsilon$ as $\frac{1}{255}$) and also clamp $\alpha$ with $0.99$ from above. Finally, \textbf{before} a Gaussian is included in the forward rasterization pass, we compute the accumulated opacity if we were to include it and stop front-to-back blending \textbf{before} it can exceed $0.9999$.}
	
	\section{Per-Scene Error Metrics}
    \label{sec:appd}
	
	\ADDITION{Tables~\ref{tab:360_scene_ssim}--\ref{tab:ttdb_scene_lpips} list the various collected error metrics for our evaluation over all considered techniques and real-world scenes. We list both the copied Mip-NeRF360 numbers and those of our runs used to generate the images in the paper; averages for these over the full Mip-NeRF360 dataset are PSNR 27.58, SSIM 0.790, and LPIPS 0.240.}

	\begin{table}[H]
	\caption{SSIM scores for Mip-NeRF360 scenes. $\dagger$ copied from original paper.}
	\scalebox{0.6}{
		\centering
	    \begin{tabular}{l|ccccc|cccc}
		~ & bicycle & flowers & garden & stump & treehill  & room & counter & kitchen & bonsai \\ \hline
		Plenoxels & 0.496 & 0.431 & 0.6063 & 0.523 & 0.509 & 0.8417 & 0.759 & 0.648 & 0.814 \\ 
		INGP-Base & 0.491 & 0.450 & 0.649 & 0.574 & 0.518  & 0.855 & 0.798 & 0.818 & 0.890 \\ 
		INGP-Big & 0.512 & 0.486 & 0.701 & 0.594 & 0.542  & 0.871 & 0.817 & 0.858 & 0.906 \\ 
		Mip-NeRF360$^\dagger$ & 0.685 & 0.583 & 0.813 & 0.744 & 0.632 & 0.913 & 0.894 & 0.920 & \textbf{0.941} \\ 
		Mip-NeRF360 & 0.685 & 0.584 & 0.809 & 0.745 & 0.631 & 0.910 & 0.892 & 0.917 & 0.938\\
		Ours-7k & 0.675 & 0.525 & 0.836 & 0.728 & 0.598  & 0.884 & 0.873 & 0.900 & 0.910 \\ 
		Ours-30k & \textbf{0.771} & \textbf{0.605} & \textbf{0.868} & \textbf{0.775} & \textbf{0.638} & \textbf{0.914} & \textbf{0.905} & \textbf{0.922} & 0.938 \\ 
	\end{tabular}
	}
	\label{tab:360_scene_ssim}
	\end{table}

	\begin{table}[H]
	\caption{PSNR scores for Mip-NeRF360 scenes. $\dagger$ copied from original paper. }
	\scalebox{0.6}{
		\centering
		\centering
		\begin{tabular}{l|ccccc|cccc}
			~ & bicycle & flowers & garden & stump & treehill  & room & counter & kitchen & bonsai \\ \hline
			Plenoxels & 21.912 & 20.097 & 23.4947 & 20.661 & 22.248 & 27.594 & 23.624 & 23.420 & 24.669 \\ 
			INGP-Base & 22.193 & 20.348 & 24.599 & 23.626 & 22.364  & 29.269 & 26.439 & 28.548 & 30.337 \\ 
			INGP-Big & 22.171 & 20.652 & 25.069 & 23.466 & 22.373  & 29.690 & 26.691 & 29.479 & 30.685 \\ 
			Mip-NeRF360$^\dagger$ & 24.37 & \textbf{21.73} & 26.98 & 26.40 & 22.87 & \textbf{31.63} & \textbf{29.55} & \textbf{32.23} & \textbf{33.46} \\
			Mip-NeRF360 & 24.305 & 21.649 & 26.875 & 26.175 & \textbf{22.929} & 31.467 & 29.447 & 31.989 & 33.397 \\
			Ours-7k & 23.604 & 20.515 & 26.245 & 25.709 & 22.085  & 28.139 & 26.705 & 28.546 & 28.850 \\ 
			Ours-30k & \textbf{25.246} & 21.520 & \textbf{27.410} & \textbf{26.550} & 22.490 & 30.632 & 28.700 & 30.317 & 31.980 \\ 
		\end{tabular}
	}
	\label{tab:360_scene_psnr}
\end{table}

	\begin{table}[H]
	\caption{LPIPS scores for Mip-NeRF360 scenes.  $\dagger$ copied from original paper.}
	\scalebox{0.6}{
		\centering
    \begin{tabular}{l|ccccc|cccc}
	~ & bicycle & flowers & garden & stump & treehill  & room & counter & kitchen & bonsai \\ \hline
	Plenoxels & 0.506 & 0.521 & 0.3864 & 0.503 & 0.540 & 0.4186 & 0.441 & 0.447 & 0.398 \\ 
	INGP-Base & 0.487 & 0.481 & 0.312 & 0.450 & 0.489  & 0.301 & 0.342 & 0.254 & 0.227 \\ 
	INGP-Big & 0.446 & 0.441 & 0.257 & 0.421 & 0.450  & 0.261 & 0.306 & 0.195 & 0.205 \\ 
	Mip-NeRF360$^\dagger$ & 0.301 & 0.344 & 0.170 & 0.261 &  0.339 & \textbf{0.211} & \textbf{0.204} & \textbf{0.127} & \textbf{0.176} \\ 
	Mip-NeRF360 & 0.305 & 0.346 & 0.171 & 0.265 & 0.347 & 0.213 & 0.207 & 0.128 & 0.179\\
	Ours-7k & 0.318 & 0.417 & 0.153 & 0.287 & 0.404  & 0.272 & 0.254 & 0.161 & 0.244 \\ 
	Ours-30k & \textbf{0.205} & \textbf{0.336} & \textbf{0.103} & \textbf{0.210} & \textbf{0.317} & 0.220 & \textbf{0.204} & 0.129 & 0.205 \\ 
\end{tabular}
	}
\end{table}

	\begin{table}[H]
	\caption{SSIM scores for Tanks\&Temples and Deep Blending scenes. }
		\centering
    \begin{tabular}{l|cc|cc}
	~ & Truck & Train & Dr Johnson & Playroom \\ \hline
	Plenoxels & 0.774 & 0.663 & 0.787 & 0.802 \\ 
	INGP-Base & 0.779 & 0.666 & 0.839 & 0.754 \\ 
	INGP-Big & 0.800 & 0.689 & 0.854 & 0.779 \\ 
	Mip-NeRF360 & 0.857 & 0.660 & \textbf{0.901} & 0.900 \\ 
	Ours-7k & 0.840 & 0.694 & 0.853 & 0.896 \\ 
	Ours-30k & \textbf{0.879} & \textbf{0.802} & 0.899 & \textbf{0.906} \\ 
\end{tabular}
\end{table}
	\begin{table}[H]
	\caption{PSNR scores for Tanks\&Temples and Deep Blending scenes. }
		\centering
		\begin{tabular}{l|cc|cc}
			~ & Truck  & Train & Dr Johnson & Playroom \\ \hline
			Plenoxels & 23.221 & 18.927 & 23.142 & 22.980 \\ 
			INGP-Base & 23.260  & 20.170 & 27.750 & 19.483 \\ 
			INGP-Big & 23.383  & 20.456 & 28.257 & 21.665 \\ 
			Mip-NeRF360 & 24.912  & 19.523 & \textbf{29.140} & 29.657 \\ 
			Ours-7k & 23.506  & 18.892 & 26.306 & 29.245 \\ 
			Ours-30k & \textbf{25.187} & \textbf{21.097} & 28.766 & \textbf{30.044} \\ 
		\end{tabular}
\end{table}

	\begin{table}[H]
	\caption{LPIPS scores for Tanks\&Temples and Deep Blending scenes. }
		\centering
    \begin{tabular}{l|cc|cc}
	~ & Truck & Train & Dr Johnson & Playroom \\ \hline
	Plenoxels & 0.335 & 0.422 & 0.521 & 0.499 \\ 
	INGP-Base & 0.274 & 0.386 & 0.381 & 0.465 \\ 
	INGP-Big & 0.249 & 0.360 & 0.352 & 0.428 \\ 
	Mip-NeRF360 & 0.159 & 0.354 & \textbf{0.237} & 0.252 \\ 
	Ours-7k & 0.209 & 0.350 & 0.343 & 0.291 \\ 
	Ours-30k & \textbf{0.148} & \textbf{0.218} & 0.244 & \textbf{0.241} \\ 
\end{tabular}
		\label{tab:ttdb_scene_lpips}
\end{table}
	
\end{appendix}

\end{document}